\documentclass[a4paper,fleqn,usenatbib]{mnras}

\newcommand\trackchanges{1}

\usepackage[T1]{fontenc}
\usepackage{ae,aecompl}

\usepackage{graphicx}    
\usepackage{amsmath}    
\usepackage{amssymb}    
\usepackage{ifthen}
\usepackage{natbib}
\usepackage{etoolbox}
\usepackage{longtable}
\usepackage{tablefootnote}

\newcommand\vs{\emph{vs.}}

\let\oldumu=\umu
\renewcommand\umu{\ifmmode\oldumu\else\math{\oldumu}\fi}

\let\oldsim=\sim
\renewcommand\sim{\ifmmode\oldsim\else\math{\oldsim}\fi}
\let\oldpm=\pm
\renewcommand\pm{\ifmmode\oldpm\else\math{\oldpm}\fi}
\newcommand\by{\ifmmode\times\else\math{\times}\fi}
\newcommand\ttt[1]{10\sp{#1}}
\newcommand\tttt[1]{\by\ttt{#1}}

\newbox{\wdbox}
\renewcommand\c{\setbox\wdbox=\hbox{,}\hspace{\wd\wdbox}}
\renewcommand\i{\setbox\wdbox=\hbox{i}\hspace{\wd\wdbox}}

\newcommand\added[1]{{\if\trackchanges1\bf\fi{#1}}}

\newcount\timect
\newcount\hourct
\newcount\minct
\newcommand\now{\timect=\time \divide\timect by 60
         \hourct=\timect \multiply\hourct by 60
         \minct=\time \advance\minct by -\hourct
         \number\timect:\ifnum \minct < 10 0\fi\number\minct}

\newcommand\mctc{\multicolumn{2}{c}}

\catcode`@=11

\newcommand\comment[1]{}

\renewcommand\math[1]{$#1$}

\let\atab=&

\let\oldmsp=\sp
\let\oldmsb=\sb
\def\sp#1{\ifmmode
           \oldmsp{#1}%
         \else\strut\raise.85ex\hbox{\scriptsize #1}\fi}
\def\sb#1{\ifmmode
           \oldmsb{#1}%
         \else\strut\raise-.54ex\hbox{\scriptsize #1}\fi}
\newbox\@sp
\newbox\@sb
\def\sbp#1#2{\ifmmode%
           \oldmsb{#1}\oldmsp{#2}%
         \else
           \setbox\@sb=\hbox{\sb{#1}}%
           \setbox\@sp=\hbox{\sp{#2}}%
           \rlap{\copy\@sb}\copy\@sp
           \ifdim \wd\@sb >\wd\@sp
             \hskip -\wd\@sp \hskip \wd\@sb
           \fi
        \fi}
\def\msp#1{\ifmmode
           \oldmsp{#1}
         \else \math{\oldmsp{#1}}\fi}
\def\msb#1{\ifmmode
           \oldmsb{#1}
         \else \math{\oldmsb{#1}}\fi}

\comment{And now to turn us totally on and off...}

\catcode`@=12

\DeclareFixedFont{\tttb}{T1}{txtt}{bx}{n}{11}  
\DeclareFixedFont{\tttm}{T1}{txtt}{m} {n}{11}  

\newcommand\chisq{\ifmmode{\chi\sp{2}}\else\math{\chi\sp{2}}\fi}
\newcommand\redchisq{\ifmmode{ \chi\sp{2}\sb{\rm red}}
                    \else\math{\chi\sp{2}\sb{\rm red}}\fi}

\newcommand\Teq{\ifmmode{T\sb{\rm eq}}\else$T$\sb{eq}\fi}
\newcommand\mjup{\ifmmode{M\sb{\rm Jup}}\else$M$\sb{Jup}\fi}
\newcommand\rjup{\ifmmode{R\sb{\rm Jup}}\else$R$\sb{Jup}\fi}
\newcommand\msun{\ifmmode{M\sb{\odot}}\else$M\sb{\odot}$\fi}
\newcommand\rsun{\ifmmode{R\sb{\odot}}\else$R\sb{\odot}$\fi}
\newcommand\mearth{\ifmmode{M\sb{\oplus}}\else$M\sb{\oplus}$\fi}
\newcommand\rearth{\ifmmode{R\sb{\oplus}}\else$R\sb{\oplus}$\fi}

\newcommand\water{H$\sb{2}$O}

\newcommand\jep{\ifmmode{\Lambda}\else\math{\Lambda}\fi}
\newcommand\lhy{\ifmmode{L\sb{\rm hy}}\else\math{L\sb{\rm hy}}\fi}
\newcommand\len{\ifmmode{L\sb{\rm en}}\else\math{L\sb{\rm en}}\fi}

\title[An overabundance of low-density Neptunes]
      {An overabundance of low-density Neptune-like planets}

\author[P. Cubillos et al.]{Patricio~Cubillos\sp{1}\thanks{E-mail: 
\href{mailto:patricio.cubillos@oeaw.ac.at}{patricio.cubillos@oeaw.ac.at}},
Nikolai~V.~Erkaev\sp{2},
Ines~Juvan\sp{1},
Luca~Fossati\sp{1},
Colin~P.~Johnstone\sp{3}, \newauthor
Helmut~Lammer\sp{1},
Monika~Lendl\sp{1},
Petra~Odert\sp{1},
and
Kristina~G.~Kislyakova\sp{1} \\
\sp{1} Space Research Institute, Austrian Academy of Sciences,
       Schmiedlstrasse 6, A-8042 Graz, Austria. \\
\sp{2} Federal Research Center "Krasnoyarsk Science Center" SB
       RAS, "Institute of Computational Modelling". \\
\sp{3} University of Vienna, Department of Astrophysics, 
       T{\"u}rkenschanzstrasse 17, 1180 Vienna, Austria}

\date{Accepted XXX. Received YYY; in original form ZZZ}

\pubyear{2016}

\begin{document}
\label{firstpage}
\pagerange{\pageref{firstpage}--\pageref{lastpage}}
\maketitle

\begin{abstract}
We present a uniform analysis of the atmospheric escape rate of
Neptune-like planets with estimated radius and mass
(restricted to $M\sb{\rm p}<30\,M_{\oplus}$).
For each planet we compute the restricted Jeans escape parameter,
$\Lambda$, for a hydrogen atom evaluated at the planetary mass,
radius, and equilibrium temperature.  Values of $\Lambda\lesssim20$
suggest extremely high mass-loss rates.
We identify 27 planets (out of 167) that are simultaneously
consistent with hydrogen-dominated atmospheres and are expected to
exhibit extreme mass-loss rates.
We further estimate the mass-loss rates ($L_{\rm hy}$) of these
planets with tailored atmospheric hydrodynamic models.  We compare
$L_{\rm hy}$ to the energy-limited (maximum-possible high-energy
driven) mass-loss rates.
We confirm that 25 planets (15\% of the sample) exhibit extremely high
mass-loss rates ($L_{\rm hy}>0.1\,M_{\oplus}{\rm Gyr}\sp{-1}$), well
in excess of the energy-limited mass-loss rates.
This constitutes a contradiction, since the hydrogen envelopes cannot
be retained given the high mass-loss rates.
We hypothesize that these planets are not truly under such high
mass-loss rates.  Instead, either hydrodynamic models overestimate the
mass-loss rates, transit-timing-variation measurements underestimate
the planetary masses, optical transit observations overestimate the
planetary radii (due to high-altitude clouds), or Neptunes have
consistently higher albedos than Jupiter planets.  We conclude that at
least one of these established estimations/techniques is consistently
producing biased values for Neptune planets.  Such an important
fraction of exoplanets with misinterpreted parameters can
significantly bias our view of populations studies, like the observed
mass--radius distribution of exoplanets for example.
\end{abstract}

\begin{keywords}
planets and satellites: atmospheres -- planets and satellites:
fundamental parameters -- hydrodynamics
\end{keywords}

\section{INTRODUCTION}

The Kepler Space Telescope mission has enabled the first estimations
of the abundance and size distribution of extrasolar planets in our
galaxy \citep[e.g.,][]{FressinEtal2013KeplerRate,
DressingCharbonneau2015apjOcurrenceHabitableMdwarfs}.  A large
fraction of these worlds have sizes intermediate between that of Earth
and Neptune, having no analog in our solar system.  Further follow-up
studies have yielded mass estimates for a large sample of Neptune-like
planets (hereafter, considered as those planets with $M\sb{\rm p} <
30 \mearth\, \approx 2 M\sb{\rm Nep}$), allowing us to study the physical properties of
exoplanets in a statistically robust manner.

The planetary radius is inferred from optical transit light-curve
observations, typically corresponding to the $\sim20-100$~mbar
level for a clear atmosphere \citep[e.g.,][]{LopezFortney2014apjMassRadius,
LammerEtal2016mnrasCorot24bAtmosphere, LecavelierEtal2008aaHD189b}.  The planetary mass, in turn,
is inferred either from measurements of the stellar radial
velocity (RV) or from transit timing variations (TTV) due to
interplanetary perturbations.
By combining the mass and radius constraints, we can infer bulk
densities and compositions.  For a given mass, the planetary radius does not
significantly depend on the interior composition, since it is made of
very incompressible materials.  On the contrary, for a given mass,
the hydrogen envelope fraction strongly correlates with the
planetary radius \citep{LopezFortney2014apjMassRadius}.

The mass--size distribution shows that a considerable fraction of
Neptune-like planets have significant envelopes (a few
percent of the core mass).  Noteworthy, a couple dozen of sub Neptunes
show bulk densities far lower than that of the
solar-system giant planets.  While it is plausible that solid cores of
a few {\mearth} can accrete significant gas envelopes before
the protoplanetary disk dissipation
\citep{LeeEtal2014apjSuperEarthAccretion,
  InamdarSchlichting2015mnrasSuperEarthFormation,
  StoeklEtal2016apjPrimordialAtmospheresAccretion,
  GinzburgEtal2016apjSuperEarthAtmospheres}, several studies find
that after the disk dispersal these gas envelopes endure significant
mass loss due to different mechanisms.

Cooling and photoevaporation, driven by the high-energy stellar
irradiation (XUV), produce moderate mass-loss rates and atmospheric
contraction.  These mechanisms act over timescales larger than $\sim\ttt{8}$~yr
\citep[e.g.,][]{LopezEtal2012apjThermalEvolution,
  OwenWu2013apjPhotoevaporation,
  GinzburgEtal2016apjSuperEarthAtmospheres}.
Hydrodynamic and photochemical processes determine the composition of
the upper atmosphere of these planets, and therefore of the escaping
particles \citet[e.g.,][]{KoskinenEtal2013icarHD209escapeI,
KoskinenEtal2013icarHD209escapeII}

Using hydrodynamic simulations,
\citet{OwenWu2016apjBoilOff} found that during the first
$\sim$Myr after the disk dispersal, low-density planets exhibit
extremely high thermally driven mass-loss rates, dubbed
``boil-off'' \citep[see also][]{LammerEtal2016mnrasCorot24bAtmosphere,
FossatiEtal2016aaAeronomicalConstraints}.  This mechanism consists
of hydrodynamic thermal evaporation (Parker wind), driven by the
internal heat from the planet, and fueled by the stellar continuum
irradiation.  During this regime, the atmosphere quickly cools and
contracts as it releases energy through the escaped particles.  In
other words, the thermal energy exceeds the gravitational energy in
atmospheric layers where the density is high enough to lead to high
mass-loss rates. Consequently, the mass-loss rate exponentially decays
until the XUV-driven photoevaporation becomes the dominant mass-loss
mechanism \citep{StoeklEtal2015aaHydroSimsEarth}.

The extremely high mass-loss rates of this regime are
unsustainable for the atmospheres of Neptune-like planets over Gyr
timescales.  Therefore, only young systems are expected to show
thermally driven mass-loss rates in excess of the XUV-driven mass-loss
rates \citep{LammerEtal2016mnrasCorot24bAtmosphere}.

Ultraviolet transit observations provide evidence of mass loss on
exoplanets \citep[e.g.,][]{Vidal-Madjar2003natHD209b,
LinskyEtal2010apjMassLossHD209b, FossatiEtal2010apjWASP12b,
BenJaffelBallester2013aaOxygenDH189b}.  For example, Ly-$\alpha$
transit observations \citep[e.g.,][]{Vidal-Madjar2003natHD209b,
KulowEtal2014apjLyAlphaGj436b,
EhrenreichEtal2015natGJ436b, BourrierEtal2016aaGJ436bEvaporation} show
transit depths much larger than in the optical.  These observations
are interpreted as absorption from
an extended region of escaping gas
beyond the Roche lobe of the planet.

However, the translation from observations into mass-loss rates is very
model-dependent.
Ly-$\alpha$ observations show large offsets between the absorption
features and the center of the line ($\pm 100$ km\,s$\sp{-1}$),
requiring additional assumptions to explain the observations.  Among
the proposed mechanisms there are:
natural broadening from large-scale confinement of material into a
higher-density region \citep{StoneProgaapj2009Winds,
OwenAdams2014mnrasMagneticMassLoss}, a combination of stellar
radiation pressure and wind interactions, where particles accelerate
by charge exchange in the wind-wind interaction region
\citep{HolmstromEtal2008natHydrogenHD209b,
TremblinChiang2013mnrasWindsChargeExchange,
KislyakovaEtal2014sciLyAlphaMagneticsHd209b,
ChristieEtal2016apjWindHydrodynamics}, or inhomogeneities of the stellar
disc at Ly-$\alpha$ light \citep{LlamaEtal2013mnrasBowSchocksHD189b}.

\subsection{Restricted Jeans Escape Parameter}
\label{sec:jeans}

\citet{FossatiEtal2016aaAeronomicalConstraints} described the
thermal escape in terms of the classical Jeans escape
parameter \citep[see e.g.,
][]{ChamberlainHunten1987PlanetaryAtmospheres}.  They generalized the
Jeans escape parameter for a hydrodynamic atmosphere subjected to the
gravitational perturbation from the host star.  They studied the upper
atmosphere, between the 100~mbar level and the Roche-lobe radius,
using hydrodynamic models.  By deriving the generalized Jeans escape
parameter across the atmospheric layers, they were able to determine
how stable a planetary atmosphere is against evaporation.

\citet{FossatiEtal2016aaAeronomicalConstraints} further defined the
restricted Jeans escape parameter:
\begin{equation}
\jep \equiv \frac{G M\sb{\rm p} m\sb{H}}{k\sb{B} \Teq R\sb{\rm p}},
\label{eq:jep}
\end{equation}
the Jeans escape parameter for a hydrogen atom evaluated at the
planetary mass ($M\sb{\rm p}$), radius ($R\sb{\rm p}$), and
equilibrium temperature ($\Teq$), where $m\sb{\rm H}$ is the mass of
the hydrogen atom, $G$ is the gravitational constant, and $k\sb{B}$ is
the Boltzmann constant.
For hydrogen-dominated atmospheres, {\jep} is an easy-to-calculate
parameter that allows one to estimate when the hydrodynamic mass-loss rate 
exceeds the XUV-driven photoevaporation.

We note that, whereas the classical Jeans escape parameter is a
variable as function of altitude in an atmosphere,
our definition of $\Lambda$ is a particular value of it
that works as a global parameter for a given
planet (no altitude dependence). This is similar to the definition
of \citet{GuillotEtal1996apjCloseGiantPlanets}, but evaluated at the
planetary equilibrium temperature.  Using this specific definition, we
empirically found the boil-off threshold at
$\Lambda\sim20$ \citep{FossatiEtal2016aaAeronomicalConstraints},
equivalent to the threshold of $R=0.1 R\sb{\rm Bondi}$
of \citet{OwenWu2016apjBoilOff}.

\citet{FossatiEtal2016aaAeronomicalConstraints} 
determined that planets with $\jep \lesssim 20$--$40$ should be
experiencing extreme mass-loss rates by comparing the 
hydrodynamic escape rate, {\lhy}, to the maximum XUV-driven
escape rate, {\len} \citep[estimated through the
energy-limited formula,][]{WatsonEtal1981icarEnergyLimited}, over a
range of {\jep} scenarios with varying stellar type, planetary mass,
and planetary radius.

In this paper we follow up the studies of atmospheric escape, modeling
and estimating the present-day mass-loss rates for a sample of known
exoplanets.  We compile a list of the exoplanets with estimated masses
and radii, and calculate their restricted Jeans escape parameter,
{\jep} (Section
\ref{sec:sample}).  Then, we compute hydrodynamic models tailored for
planets suspected of being in the boil-off regime (Section
\ref{sec:hydro}).  Finally we discuss the observational and physical
implications of our findings (Section \ref{sec:discussion}).

\section{SAMPLE OF KNOWN NEPTUNES}
\label{sec:sample}

We compiled our sample by collecting and crosschecking the lists of
exoplanets from the Nasa Exoplanet
Archive\footnote{\href{http://exoplanetarchive.ipac.caltech.edu/}
{http://exoplanetarchive.ipac.caltech.edu/}} the Exoplanets Data
Explorer\footnote{\href{http://exoplanets.org/}
{http://exoplanets.org/}} \citep{HanEtal2014paspExoplanetsDotOrg}, and The Extrasolar Planets
Encyclopaedia\footnote{\href{http://exoplanet.eu/}
{http://exoplanet.eu/}} \citep{SchneiderEtal2012aspcExoplanetDotEu}. We selected the targets with measured
planetary radii and masses, whose mass is less than $\sim2$~Neptune
masses ($M\sb{\rm p} < 30~M\sb{\oplus}$).  We adopted stellar
rotational angular velocity ($\Omega\sb{\rm rot}$)
from \citet{McQuillanEtal2013apjStellarRotations} and ages
from \citet{Morton2016apj1284KOIvalidated}.  Our final sample consists
of 167 planets (Table \ref{table:allplanets}).

\begin{figure*}
\includegraphics[width=\linewidth, clip]{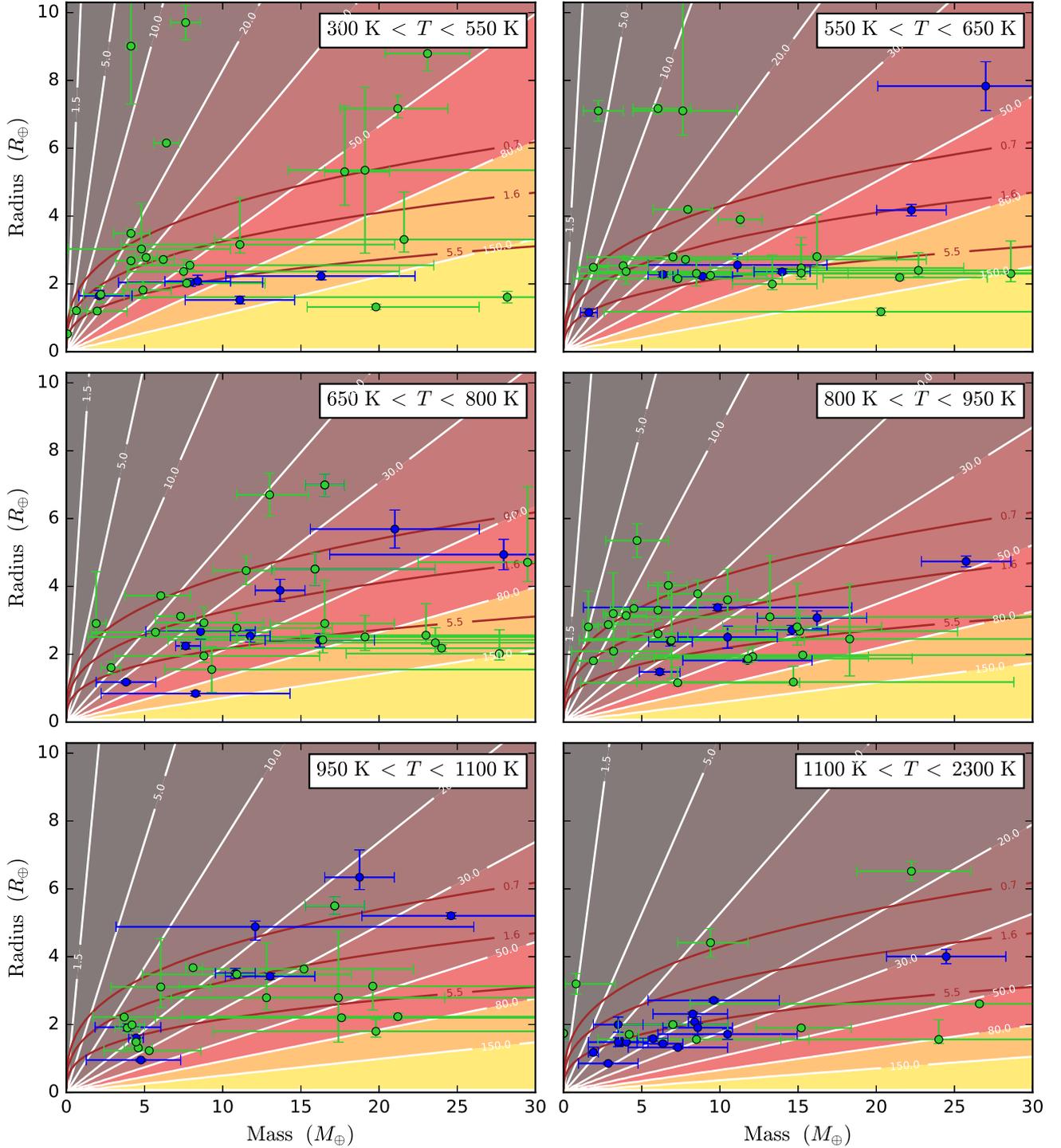}
\caption{Mass--radius--{\Teq} distribution of the known Neptune-like
  planets.  The blue and green markers denote RV- and TTV-measured
  masses, respectively.  From top to bottom, the brown solid lines
  denote constant-density curves corresponding to Saturn's, Neptune's,
  and Earth's bulk density in g\,cm$\sp{-3}$, respectively.  The
  background color contour denotes the value of {\jep} for the mean
  equilibrium temperature of the planets in each panel: 455, 600, 715,
  870, 1025, and 1440 K.  Note that the {\jep}-contours work only as a
  guideline, Table \ref{table:allplanets} gives the specific values
  for each planet.}
\label{fig:allplanets}
\end{figure*}

Fig.\ \ref{fig:allplanets} shows the mass--radius--{\Teq}
distribution for the planets in this sample.  We calculated the
equilibrium temperature assuming zero Bond albedo and
efficient day--night energy redistribution.  We split this figure into
temperature bins to show representative {\jep} contours for the
planets.

This is a heterogeneous sample of planets.  The large majority of
these planets (90\% of the sample) were discovered using the transit
method from the Kepler Space Telescope.  The masses of these planets
were estimated using the TTV and RV methods (70\% and 30\% of the
sample, respectively),
with only a few systems having both TTV and RV constraints: Kepler-18,
Kepler-89, K2-19, and WASP-47. 

The distribution of planets in Fig.\ \ref{fig:allplanets} reflects the
selection biases from each observing method.
Most RV planets fall into the two panels with higher {\Teq}
because the RV method favors planets orbiting closer to
their host stars.
Furthermore, for a given planet size, the RV method tends to find
planets with higher mass while the sensitivity of TTVs is more
uniform \citep{WeissMarcy2014apjlMassRadius, Steffen2016mnrasRVvsTTV}.
\citet{Jontof-HutterEtal2014apjTTVkepler79} argued that larger planets
have deeper transits yielding more precise transit times.

Table \ref{table:allplanets} lists the observed and derived parameters
for each planet.
We selected the planets consistent with an envelope mass
fraction larger than 1\% (88 planets) by comparing the observed radii
to the models of \citet{LopezFortney2014apjMassRadius}, for the given
planetary masses, stellar ages, and incident fluxes.
From this sub sample, we identify 28 planets (17\% of the sample) with $\jep <
20$, and thus should exhibit extreme mass-loss rates.
We note that all of these systems are old ($>$1 Gyr).  We observe some
exceptional cases in system parameters, extremely low-bulk-density
planets ($\rho\sb{\rm p} < 0.2$\,g\,cm$\sp{-3}$) like Kepler-33,
Kepler-51, Kepler-79, Kepler-87, or Kepler-177; and extremely unstable
atmospheres ($\jep < 5$) like Kepler-33~c or Kepler-51~b.

\section{HYDRODYNAMIC RUNS}
\label{sec:hydro}

In this section we present hydrodynamical models of the upper
atmosphere of selected Neptune-like planets, which allow us to directly
estimate the present-day hydrodynamic mass-loss rate.  We further determine
whether their mass-loss rate is higher than the upper-limit XUV-driven
mass-loss rate.
We covered a range of {\jep} from 0 to 50, including all
hydrogen-dominated atmospheres with $\jep < 20$, some
higher-density planets, and some planets with larger {\jep}.

To compute {\lhy}, we applied the one-dimensional upper-atmosphere
hydrodynamic model of \citet{ErkaevEtal2016mnrasThermalLoss}.
This model solves the hydrodynamic system of equations for mass,
momentum, and energy conservation, considering the absorption of
stellar XUV flux and accounting for the particles' ionization,
dissociation, recombination, and Ly-$\alpha$ cooling\footnote{See the appendix
in \citet{ErkaevEtal2016mnrasThermalLoss} for a detailed
description.}.

For the XUV absorption we assume a spherically symmetric distribution
of density, deviations from this symmetry do not seem to be essential.
One could expect that difference between 1D and 3D models would be
more pronounced for cases of larger Jeans escape parameter, when the
hydrodynamic flow is driven mainly by the XUV flux, which may have strong
asymmetries.  But in such case the mass-loss rate is rather close to the
energy-limited escape value, as observed in existing 2D \citep{KhodachenkoEtal2015apjXUVmassLoss} and 3D \citep{TripathiEtal2015apj3Dphotoevaporation} simulations.

For hot Jupiters, our model produces similar mass-loss rates as most
other hydrodynamic models \citep{ErkaevEtal2016mnrasThermalLoss}.
For the Neptune-like planet GJ~436\,b, our model predicts a mass-loss
rate of $2${$\times$}$\ttt{-9}$ g\,s$\sp{-1}$, consistent with the
values from \citet{EhrenreichEtal2015natGJ436b}.

As in \citet{TuEtal2015aaStellarEvolutionaryTracks} and
\citet{JohnstoneEtal2015apjStellarRotationEvolution}, we calculate the
stellar X-ray and EUV luminosities using the scaling laws of
\citet{WrightEtal2011apjStellarActivityRotation} to convert the
stellar rotation rates and masses into X-ray luminosities, and
\citet{SanzForcadaEtal2011aaXUVevaporation} to convert the X-ray
luminosities into EUV luminosities.  When the stellar rotation rate is
not known, we use the gyrochronological relation of
\citet{MamajekHillenbrand2008apjAgeActivityRotation} to convert
stellar age into rotation rate.

We calculate the upper-limit XUV-driven escape using the energy-limited formula
\citep{WatsonEtal1981icarEnergyLimited,
  ErkaevEtal2007aapHotJupitersRocheLobe}:
\begin{equation}
L\sb{\rm en} = \frac{\pi \eta R\sb{\rm p} (R\sb{\rm XUV}\sp{\rm eff})\sp{2} F\sb{\rm XUV}} {G M\sb{\rm p} m\sb{H} K},
\end{equation}
with $\eta = 15$\% the net heating
efficiency \citep{ShematovichEtal2014aaHeatingEfficiency}, $R\sb{\rm
XUV}\sp{\rm eff}$ the radius where the bulk of the XUV energy is
absorbed, and $K$ the potential
energy reduction factor due to stellar tides, which depends on the
Roche-lobe boundary radius.

It is important to remark that both the stellar flux and the planetary
parameters vary over their respective lifespan, inducing a variation
in the planetary mass-loss rate with time \citep[see,
e.g.,][]{LecavelierEtal2004aaAtmosphericEscape,
Lecavelier2007aaExoplanetsEvaporation}.  In particular, for the
boil-off regime, \citet{OwenWu2016apjBoilOff} argue that the mass-loss
rate is exponentially sensitive to the planetary radius at small
$\Lambda$.  Thus, during boil-off, strong cooling and contraction
quickly reduce $R\sb{\rm p}$ (and hence $\lhy$) in timescales of $\sim
1$\,Myr until $\Lambda \sim 20$.
The XUV stellar fluxes we compute take into account the age of the
star, whereas the planetary radii and masses are taken directly from
the reported values.  Therefore, the mass-loss rates we compute should
be considered as a present-day snapshot of their values.

\begin{figure}
\includegraphics[width=\linewidth, clip]{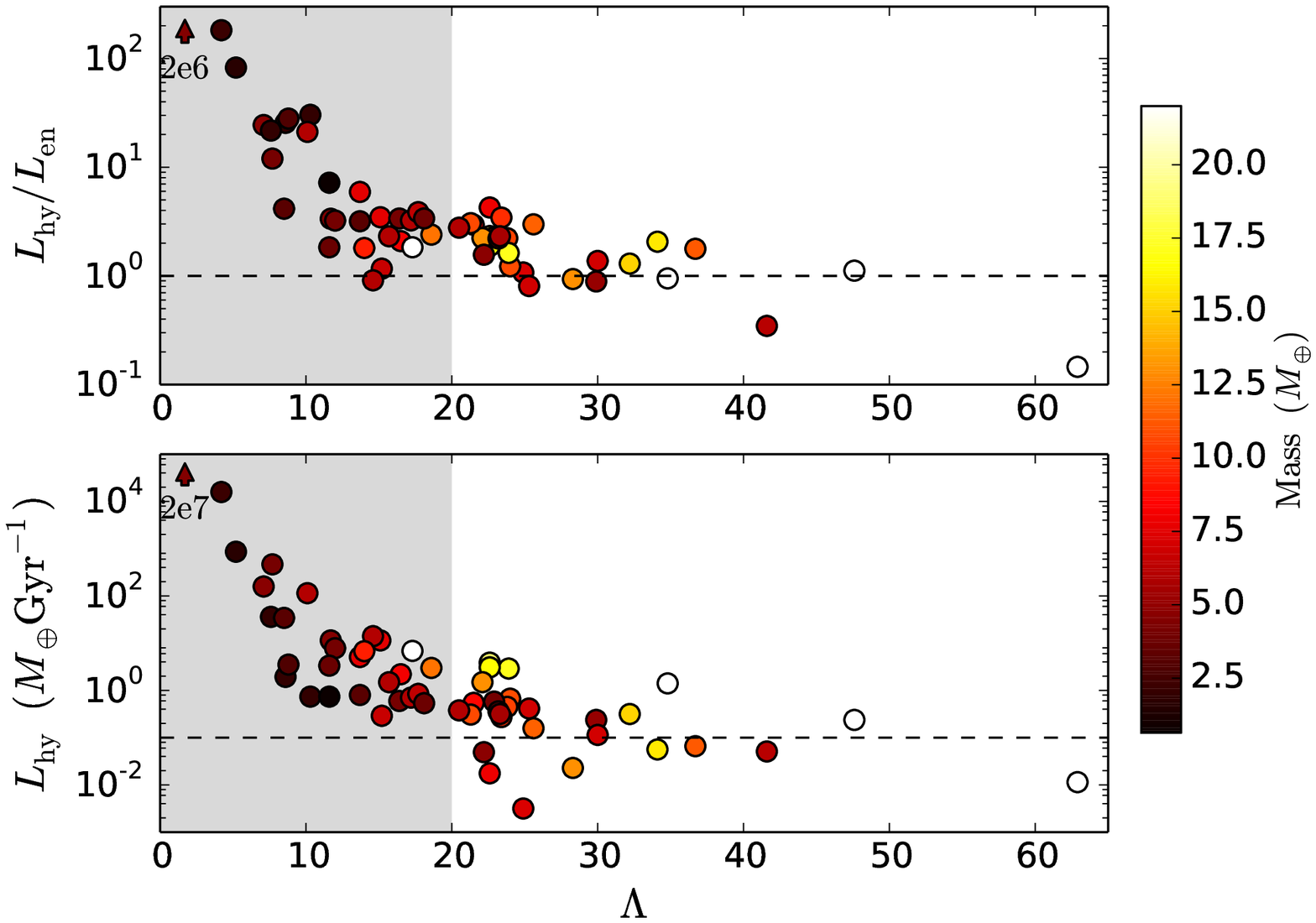}
\caption{Top: Ratio between the hydrodynamic and energy-limited
  mass-loss rate {\vs}\ {\jep}.  The color scale denotes the planetary
  mass.  The background color marks the regions with dominant extreme
  thermal mass-loss rate (gray) and XUV-driven photoevaporation
  (white). The labeled triangle denotes the mass-loss values for the
  extreme case of Kepler-33~c.  Bottom: Hydrodynamic mass-loss
  rate {\vs}\ {\jep}.}
\label{fig:correlations}
\end{figure}

Table \ref{table:allplanets} and Fig.\ \ref{fig:correlations} show the
hydrodynamic and energy-limited mass-loss rates for each modeled
planet.
Note that the estimated {\lhy} is the total mass-loss rate, including
photoevaporation.  Thus, we can infer whether photoevaporation
(${\lhy}\approx {\len}$) or boil-off (${\lhy} \gg {\len}$) is
the dominant mass-loss mechanism.
Given the variability of {\lhy} and {\len} with the adopted
model parameters ($F\sb{\rm XUV}$ and $\eta$), we consider that
planets with $\lhy/\len > 2.5$ are in the boil-off regime
\citep{LammerEtal2016mnrasCorot24bAtmosphere}.

We confirm that {\jep} is a good indicator of extreme mass-loss
rates (Fig.\ \ref{fig:correlations}, top panel). For $\jep < 15$,
{\lhy/\len} is strongly correlated with {\jep}, exponentially
increasing with decreasing {\jep}.  In the intermediate range $15
< \jep < 25$, there is a transition from boil off to
photoevaporation-dominated mass-loss rate.  Then, for $\jep > 25$,
{\lhy/\len} remains constant as the mass-loss rate is dominated by the
XUV-driven photoevaporation.  We find no significant correlation of
{\jep} with other observed parameters.
Furthermore, 19 of the 27 previously identified planets
with hydrogen-dominated atmospheres and $\jep < 20$ satisfy $\lhy/\len
> 2.5$.  Six additional low-density planets with
$\jep \leq 25$ also present an excess mass-loss rate.
The bottom panel of Fig.\ \ref{fig:correlations} shows the computed
hydrodynamic mass-loss rates.  All planets with excess
mass-loss rates ($\lhy/\len > 2.5$) show hydrodynamic mass-loss rates above
$0.1\ \mearth{\rm Gyr}\sp{-1}$.  At this rate, these planets should
have already lost their hydrogen atmospheres.

The scatter in {\lhy} (Fig.\ \ref{fig:correlations}, bottom panel)
seems to be mainly correlated with the planetary mass: at a given
{\jep} more massive planets have higher mass-loss rates.
On the contrary, the main source for the scatter in {\lhy/\len} (Fig.\
\ref{fig:correlations}, top panel) is unclear; these variations seem to
be a more complex combination of several factors, e.g., Roche-lobe
radius, incident stellar flux, XUV absorption radius, etc.

\section{DISCUSSION}
\label{sec:discussion}

The observed low bulk densities of many Neptune-like planets require a
significant hydrogen envelope fraction.
However, their estimated restricted Jeans escape parameter, {\jep}, and
further hydrodynamic modeling indicate that 25 planets in our sample
(15\%) should be experiencing extremely high mass-loss
rates ($\lhy > 0.1\ \mearth{\rm Gyr}\sp{-1}$) in excess of the maximum
XUV-driven photoevaporation (${\lhy/\len} > 2.5$).

Considering the age of the systems ($\gtrsim$ Gyr), it is improbable
that these Neptunes have retained their hydrogen envelopes
until the present day.  In addition, the short timescale of the
boil-off regime ($\lesssim$ Myr) makes it unlikely that we
are observing a transient phenomenon by chance.
This contradiction leads us to consider biases on the mass-loss model
estimations, the measured physical parameters, or the physical
interpretation of the observations.

\subsection{Hydrodynamic-model Bias?}
\label{sec:massbias}

Validating model mass-loss rates is challenging because to date there
are no direct observational measurements.
UV transit observations depend as much on the hydrodynamic models
inside the planetary Roche lobe as in the particle dynamics after.
This would require a treatment of a number of physical processes that
are largely unconstrained (e.g., magnetic fields, stellar radiation
pressure, wind-wind interaction).

The heating efficiency rate $\eta$ is one of the least constrained
parameters in our model, for which we assume a value of $0.15$.
A detailed calculation of $\eta$ is complicated, as it varies with
altitude and it requires a kinetic approach considering many chemical
reactions.  \citet{ShematovichEtal2014aaHeatingEfficiency} estimated
values of $\eta$ between 0.1 and 0.2 for hot Jupiters.  Assuming
larger heating efficiency one can increase the mass-loss rates;
however, \citet{OwenJackson2012mnrasXUVevaporation} discarded values
larger than $\eta=0.4$.  \citet{SalzEtal2016aaEnergyLimited} found
little variation of the heating efficiency ($\eta \approx 0.1$--$0.3$)
with gravitational potential, except for the most compact and massive
planets.

One process that we do not consider is magnetic
fields.  \citet{OwenAdams2014mnrasMagneticMassLoss}
and \citet{KhodachenkoEtal2015apjXUVmassLoss} found that magnetic
fields can suppress mass-loss rates by approximately an order of
magnitude by confining the outflowing material into smaller opening
angles around the poles.

\subsection{Planetary-mass Bias?}
\label{sec:massbias}

One explanation to this contradiction is that TTV analyses are
underestimating the planetary masses, a possibility already considered
by \citet{WeissMarcy2014apjlMassRadius}.  A more massive planet would
have a stronger gravitational pull, decreasing the mass-loss rate.
\citet{Steffen2016mnrasRVvsTTV} argues that
the mass measurements derived from RV and TTV are comparably reliable
since the physics behind TTV measurements (gravity) is well
understood.  However, given the complexity and rapid-pace development
of exoplanet RV and TTV analyses, as data-reduction techniques improve,
their mass estimations continuously get overturned or refined.
For example, TTV-estimated masses for
Kepler-114\,c \citep{Xie2014apjsTTV},
Kepler-231\,c \citep{KippingEtal2014apjTTVkepler231}, and
Kepler-56\,b \citep{HuberEtal2013sciTTVkepler56} differ by a
factor of ten from a nearly contemporaneous
estimation \citep{HaddenLithwick2014apj139KeplerTTV}.
Cases of extremely high densities \citep[e.g., Kepler-327\,c,
$\rho\sb{\rm p} \approx 70$
g\,cm$\sp{-3}$][]{HaddenLithwick2014apj139KeplerTTV} or even negative
RV-estimated masses \citep[e.g.,][]{MarcyEtal2014apjsRVkeplers} serve
as reminders to use the data-reduction and statistical tools with
caution.  One should incorporate all available prior information
(e.g., in a Bayesian posterior sampling) to avoid results at odds with
physical principles.

In our sample, with the exception of Kepler-94~b, all
boil-off planets have TTV-determined masses.  Ideally,
comparing the TTV and RV results for a given system would test the
reliability of these techniques.  However, due to the observing
limitations of each method, only a couple of systems have allowed for
independent TTV and RV analyses.  For example, for the Kepler-89
system, \citet{MasudaEtal2013apjTTVkepler89} found discrepancies
between the TTV and RV-mass estimations.
On the contrary, for the Kepler-18 and WASP-47 systems, TTV and RV
analyses return consistent
results \citep[][]{CochranEtal2011apjsTTV+RVkepler18,
DaiEtal2015apjWASP47RV}.

If we adopted masses such that $\jep \equiv 20$ ($M\sb{20}$,
keeping all other parameters fixed), 13 of the 19 low-density
extreme-mass-loss planets with $\jep < 20.0$ would have a mass correction
larger than 1$\sigma\sb{\rm M}$, with a mean of 5.0$\sigma\sb{\rm M}$
(see Table \ref{table:boilplanets}).

{\renewcommand{\arraystretch}{1.21}
\begin{table*}
\caption{Low-density boil-off planet parameters. {\jep}, $M\sb{\rm
    p}$, $R\sb{\rm p}$, and {\Teq} are as in Table
  \ref{table:allplanets}.
  $M\sb{20}$, $R\sb{20}$, and $A\sb{20}$ are the modified mass,
  radius, and bond albedo such that $\jep=20$. $p\sb{\rm phot}$ is the
  photospheric pressure corresponding to $R\sb{\rm p}$ when we adopt
  $R\sb{20}$ as the 100~mbar pressure level.}
\label{table:boilplanets}
\begin{tabular}{@{\extracolsep{\fill}}lr@{\hspace{0.7cm}}r@{}lr@{\hspace{1cm}}r@{}lrc@{\hspace{1cm}}rr}
\hline
Planet name & \multicolumn{1}{c}{\jep} & \mctc{$M\sb{\rm p}$} & $M\sb{20}$ & \mctc{$R\sb{\rm p}$} & $R\sb{20}$ & ${p\sb{\rm phot}}$ & \Teq & ${A\sb{20}}$ \\
            &                          & \mctc{\mearth}       & \mearth    & \mctc{\rearth}       & \rearth    & bar                    & \multicolumn{1}{c}{K} &     \\
\hline
Kepler-11 c  &   8.8 &   2.86 & $\sb{ -1.59}\sp{ +2.86}$ &  6.5  &   2.87 & $\sb{ -0.06}\sp{ +0.04}$ &  1.26 & $6.8\tttt{-13}$  &  859 & 0.96 \\
Kepler-11 f  &  10.3 &   1.91 & $\pm 0.95$               &  3.7  &   2.49 & $\sb{ -0.07}\sp{ +0.04}$ &  1.29 & $2.4\tttt{-11}$  &  562 & 0.93 \\
Kepler-114 d &  18.1 &   3.81 & $\pm 1.59$               &  4.2  &   2.54 & $\pm 0.28$               &  2.30 & $1.4\tttt{-03}$  &  628 & 0.32 \\
Kepler-177 b &   7.6 &   1.91 & $\sb{ -0.32}\sp{ +0.64}$ &  5.0  &   2.91 & $\sb{ -0.30}\sp{ +1.53}$ &  1.10 & $4.7\tttt{-14}$  &  655 & 0.98 \\
Kepler-177 c &  13.7 &   7.63 & $\sb{ -3.18}\sp{ +3.50}$ & 11.1  &   7.10 & $\sb{ -0.72}\sp{ +3.71}$ &  4.87 & $5.5\tttt{-08}$  &  594 & 0.78 \\
Kepler-254 c &  13.7 &   3.20 & $\sb{ -2.70}\sp{ +3.70}$ &  4.7  &   2.09 & $\sb{ -0.16}\sp{ +0.84}$ &  1.44 & $5.5\tttt{-08}$  &  845 & 0.78 \\
Kepler-29 b  &  11.7 &   4.50 & $\pm 1.50$               &  7.7  &   3.35 & $\pm 0.22$               &  1.95 & $5.0\tttt{-10}$  &  872 & 0.88 \\
Kepler-29 c  &  12.0 &   4.00 & $\pm 1.30$               &  6.6  &   3.14 & $\pm 0.20$               &  1.89 & $1.2\tttt{-09}$  &  802 & 0.87 \\
Kepler-305 c &  17.2 &   6.04 & $\sb{ -2.22}\sp{ +2.54}$ &  7.0  &   3.30 & $\sb{ -0.33}\sp{ +0.82}$ &  2.83 & $1.4\tttt{-04}$  &  808 & 0.46 \\
Kepler-307 b &   8.5 &   3.18 & $\pm 0.64$               &  7.4  &   3.20 & $\sb{ -0.46}\sp{ +1.20}$ &  1.36 & $4.1\tttt{-13}$  &  883 & 0.97 \\
Kepler-307 c &   5.2 &   1.59 & $\sb{ -0.32}\sp{ +0.64}$ &  6.1  &   2.81 & $\sb{ -0.42}\sp{ +1.05}$ &  0.74 & $2.3\tttt{-16}$  &  818 & 0.99 \\
Kepler-33 c  &   1.7 &   0.80 & $\sb{ -0.70}\sp{ +2.50}$ &  9.4  &   3.20 & $\pm 0.30$               &  0.27 & $6.7\tttt{-20}$  & 1113 & 0.99 \\
Kepler-33 d  &   7.1 &   4.70 & $\pm 2.00$               & 13.3  &   5.35 & $\pm 0.49$               &  1.89 & $1.3\tttt{-14}$  &  941 & 0.98 \\
Kepler-51 b  &   4.2 &   2.22 & $\sb{ -0.95}\sp{ +1.59}$ & 10.5  &   7.10 & $\pm 0.30$               &  1.50 & $2.3\tttt{-17}$  &  561 & 0.99 \\
Kepler-51 c  &   7.7 &   4.13 & $\pm 0.32$               & 10.8  &   9.01 & $\sb{ -1.71}\sp{ +2.81}$ &  3.45 & $5.3\tttt{-14}$  &  453 & 0.98 \\
Kepler-51 d  &  15.1 &   7.63 & $\pm 0.95$               & 10.1  &   9.71 & $\pm 0.50$               &  7.34 & $1.3\tttt{-06}$  &  394 & 0.67 \\
Kepler-79 d  &  10.1 &   6.04 & $\sb{ -1.59}\sp{ +2.10}$ & 12.0  &   7.17 & $\sb{ -0.16}\sp{ +0.13}$ &  3.61 & $1.4\tttt{-11}$  &  633 & 0.94 \\
Kepler-79 e  &  16.4 &   4.13 & $\sb{ -1.11}\sp{ +1.21}$ &  5.0  &   3.49 & $\pm 0.13$               &  2.87 & $2.7\tttt{-05}$  &  546 & 0.54 \\
Kepler-87 c  &  17.7 &   6.40 & $\pm 0.80$               &  7.2  &   6.15 & $\pm 0.09$               &  5.46 & $5.1\tttt{-04}$  &  444 & 0.38 \\
\hline
\end{tabular}
\end{table*}
\normalsize
}

\subsection{Planetary-radius Bias?}
\label{sec:radiusbias}

Another possibility is that we are misinterpreting the observed
transit radius.  For clear atmospheres in Neptune-like
planets, the observed optical transit radius corresponds to the
$\sim$20--100~mbar level \citep{LopezFortney2014apjMassRadius,
LammerEtal2016mnrasCorot24bAtmosphere}.  However, if a planet has
high-altitude, optically thick clouds/hazes, the transit radius would
be overestimating the planetary radius.  If this is the case, then the
true 100\,mbar altitude would correspond to a smaller radius than the
observed radius, yielding moderate mass-loss rates.  Spectroscopic
analyses already indicate that many Neptune and sub-Neptune sized
planets show flat transmission spectra, consistent with gray opacity
hazes or clouds \citep[e.g.,][]{KreidbergEtal2014natCloudsGJ1214b,
KnutsonEtal2014natFlatGJ436b, EhrenreichEtal2014aaHubbleGJ3470b}.

If we adopted radii such that $\jep \equiv 20$ ($R\sb{20}$, keeping
all other parameters fixed), 18 of the 19 low-density boil-off planets
with $\jep<20.0$ would have a radius correction larger than
1$\sigma\sb{\rm R}$, with a mean of 8.5$\sigma\sb{\rm R}$ (see
Table \ref{table:boilplanets}).
Adopting $R\sb{20}$ as the 100~mbar pressure level, we computed
hydrostatic-equilibrium pressure--radius profiles to calculate the
observed transit photospheric pressure, $p\sb{\rm phot} = p(R\sb{\rm
p})$.  Table \ref{table:boilplanets} shows that a third of the planets
could have photospheres at pressures between $\ttt{-3}$ bar and $\ttt{-7}$
bar, compatible with the location of cloud decks on exoplanets.

\begin{figure}
\includegraphics[width=\linewidth, clip]{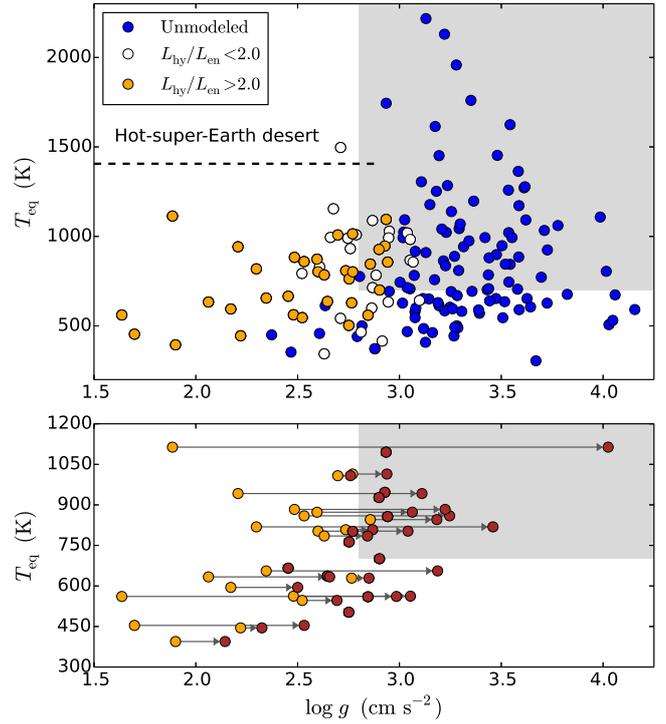}
\caption{Planetary surface gravity {\vs}\ equilibrium-temperature.
  The orange, white, and blue circles (top panel) denote the extreme
  mass-loss rate, moderate mass-loss rate, and unmodeled planets,
  respectively.  The gray and white background colors denote the
  region where planets have clear and cloudy-atmospheres,
  respectively, derived by \citet{Stevenson2016apjCloudsProminece} for
  hot Jupiters.  The horizontal dashed line denotes the lowest
  incident-stellar-flux boundary (650 $F\sb{\oplus}$) of the
  hot-super-Earth
  desert \citep{LundkvistEtal2016natHotSuperEarthDesert}.  The bottom
  panel shows the expected $\log g$ of the boil-off planets if their
  radii were corrected to have $\jep \equiv 20.0$ (brown
  circles) relative to their initial position (orange circles).}
\label{fig:gteq}
\end{figure}

When we plot the planetary surface gravity, $\log g$, {\vs} equilibrium
temperature for the planets in our sample (Fig.\ \ref{fig:gteq}, top
panel), we see that all boil-off planets are grouped in the
$\log g < 3.0$ region.
While it is not surprising that these planets are concentrated at low
$\log g$ values (given the similarity in mass and radius dependency of
$\log g\propto M\sb{\rm p}\,R\sb{\rm p}\sp{-2}$ and $\jep\propto
M\sb{\rm p}\,R\sb{\rm p}\sp{-1}$), it is interesting that this region
limits closely to the boundary beweeen strong and weak J-band {\water}
features observed in hot Jupiters, associated to cloudy {\vs} clear
atmospheres, respectively \citep{Stevenson2016apjCloudsProminece}.
Regardless of whether there is a link between cloud processes on hot
Jupiters and Neptune-like planets ---two completely different
samples---, this graph helps to identify planets with a higher
prominence of clouds/hazes.  Note that the derived surface
gravity of these planets is also affected by the
observational biases.  Fig.\ \ref{fig:gteq} (bottom panel) shows the
corrected $\log g$--{\Teq} position of the boil-off planets if we
assumed $R\sb{20}$ as the 100~mbar radius.

An important remark to highlight is that, while our analysis suggests
that boil-off planets can have instead moderate mass-loss rates if
they have cloudy atmospheres, we cannot claim that the remaining
planets have clear atmospheres.  They can either host clear or cloudy
atmospheres.  This occurs because our analysis is based on the
dominant effect of hydrogen in the observed planetary radius.  There
can be cloudy planets with secondary atmospheres, or even
hydrogen-dominated atmospheres hosting lower-altitude clouds that do
not alter much the observed transit radius.

\subsection{Bond-albedo Bias?}
\label{sec:radiusbias}

The cloudy-atmospheres scenario has a second consequence on the
properties of these planets.  The high-altitude cloud/haze layer would
reflect a larger fraction of the incident stellar irradiation,
decreasing the amount of energy deposited in the atmosphere, and
consequently decreasing the estimated mass-loss rates.  Low-density
sub Neptunes (on average) would then differ from hot Jupiters.
Studies of the aggregate sample of hot Jupiters show low geometric
optical albedos ($A\sb{g} \approx 0.1$) and somewhat higher Bond
albedos \citep[$A\sb{B} = 0.4 \pm
0.1$,][]{SchwartzCowan2015mnrasAlbedos}, suggesting that hot Jupiters
have a layer of optical absorbers above an infra-red reflective cloud
deck.  Our cloudy-atmosphere case aligns well with the higher
geometric albedos ($A\sb{g} \approx 0.16-0.3$) found on super-Earth
planets \citep[$R\sb{\rm p} <
2.5 \rearth$,][]{Demory2014apjKeplerSuperEarthAlbedos}.

Table \ref{table:boilplanets} shows the required values of the
Bond albedo such that $\jep \equiv 20.0$ ($A\sb{20}$, keeping all
other parameters fixed).  Most of the planets in this list would
require Bond albedos higher than any observed value from solar or
extrasolar gas-giant planet.

If we confirm the cloudy nature of these atmospheres, this analysis
places strong constraints on the atmospheric properties of the
ensemble of low-density sub Neptunes.  These planets would present
high-altitude opaque clouds/hazes that increase the observed radii.
Observationally, the value of their restricted Jeans escape parameter
{\jep}, and also their position in the $\log g$--{\Teq} plane, would
be simple and direct indicators of the cloud/haze prominence on the
atmospheres of these planets.

On a different but still interesting topic, the distribution of
planets in the $\log g$--{\Teq} plane (Fig.\ \ref{fig:gteq} top panel)
agrees well with the proposed hot-super-Earth
desert \citep{LundkvistEtal2016natHotSuperEarthDesert}.  The boundary
at 650 times the incident flux on Earth ($F\sb{\oplus}$) directly
translates into an equilibrium temperature of $\sim1400$ K.  From the
planets with known mass and radius, we observe a dearth of planets
above this line.  Unlike the $R\sb{\rm p}$--$F\sb{\oplus}$ figure
from \citet{LundkvistEtal2016natHotSuperEarthDesert}, this figure
further constrains the parameter space of the planetary physical
properties of the desert by adding information from the masses
(through the surface gravity).

Furthermore, by considering the boil-off mass-loss regime, our results
(Fig.\ \ref{fig:gteq} bottom panel) suggest that the exoplanet desert
extends even further.  Be it the mass or the radius of the planets
being misinterpreted, boil-off planets should have a higher surface
gravity, further depleting a region below the $\Teq=1400$ K boundary,
down to $\sim600$ K.

\section{CONCLUSIONS}
\label{sec:conclusion}

We studied the mass-loss rates of the sample of 167 confirmed planets
with masses $M\sb{\rm p} < 30 \mearth$.  For 15\% of the planets in
this sample, we found that their planetary mass, radius, and
equilibrium temperature lead to paradoxical results.  Their
observed low bulk densities imply hydrogen-dominated atmospheres.
However, hydrodynamical models indicate extremely high mass-loss rates
($\lhy > 0.1\ \mearth{\rm Gyr}\sp{-1}$), thus being unable to retain
their hydrogen envelopes (boil-off regime).

Our hydrodynamical models' mass-loss rates agree well with
state-of-the-art models from the literature.  Unfortunately, current
observations do not offer a direct, model-independent validation of
escape rates.
Magnetic fields (however largely unknown) could offer a solution of
this paradox by decreasing the predicted mass-loss
rates \citep{OwenAdams2014mnrasMagneticMassLoss,
KhodachenkoEtal2015apjXUVmassLoss}.  Alternatively, the boil-off
timescales of $\sim\ttt{6}$ yr \citep{OwenWu2016apjBoilOff} could have
been grossly underestimated.  We need more detailed studies of the
mass-loss time evolution during the boil-off regime.

If the model mass-loss estimations are correct,
we hypothesize that either the TTV studies are underestimating the
planetary masses or that these planets have high-altitude clouds/hazes
producing overestimated transit radii, high albedos, or both.
The cloudy/hazy scenario is further supported by the higher albedo
found for super Earths, and a similar correlation as that between
$\log g$--{\Teq} and clear/cloudy atmospheres for hot Jupiters.

Whether we are misinterpreting their observed transit radii
or underestimating their TTV masses, a fraction of $\sim$10--20\%
Neptunes with biased parameters will have a large impact on the
population studies of the mass--radius relationship.
Both empirical laws and physically motivated models tracing the
observed mass-radius distribution of Neptunes would be disrupted by
this sub-sample, simply because the observed values are not
representative of the underlying properties of exoplanets.  This could
explain, for example, why studies like \citet{WolfgangLopez2015apjKeplerMassRadius}
do not find a direct mass--radius relationship for sub Neptune
planets.
In the future, population studies should consider the added
complexity introduced by this rather large sub-sample of planets.

Conclusively solving this puzzle involves a multitude of studies, from
the confirmation of low-densities with RV measurements, to the
consistent modeling of high-altitude clouds/hazes.  This study
generates many open questions for the future, for example, can
clouds/hazes form and remain at such high altitudes and under strong
stellar irradiation conditions? Or how do reflective clouds affect the
atmospheric albedo, and hence mass loss?

\section*{ACKNOWLEDGMENTS}

We thank contributors to SciPy,
Matplotlib \citep{Hunter2007ieeeMatplotlib}, the Python Programming
Language, and contributors to the free and open-source community.  We
thank the anonymous referee for comments that significantly improved
the quality of the paper. This research has made use of the Exoplanet
Orbit Database and the Exoplanet Data Explorer at exoplanets.org and
the NASA Exoplanet Archive, which is operated by the California
Institute of Technology, under contract with the National Aeronautics
and Space Administration under the Exoplanet Exploration Program.  We
acknowledge the Austrian Forschungsf{\"o}rderungsgesellschaft FFG
projects ``RASEN'' P847963 and ``TAPAS4CHEOPS'' P853993, the Austrian
Science Fund (FWF) NFN projects S11607-N16 and S11604-N16, and the FWF
project P27256-N27. N.V.E. acknowledges support by the RFBR grant
No. 15-05-00879-a and No. 16-52-14006 ANF-a.

\bibliographystyle{mnras}

\appendix

\section{PARAMETERS OF THE KNOWN NEPTUNE PLANETS}

Table \ref{table:allplanets} list the observed and derived
parameters for the known sample of Neptune-like planets.

{\renewcommand{\arraystretch}{1.21}
\setlength\tabcolsep{0.1cm}

\begin{table*}
\caption{
Observed and derived parameters for the sub-Neptune planet sample.
This table only includes planets with estimated mass ($M\sb{\rm p}$)
and transit radius ($R\sb{\rm p}$), with $M\sb{\rm p} < 30 \mearth$.
The equilibrium temperature (\Teq) assumes zero Bond albedo and
efficient day--night energy redistribution.
The restricted Jeans escape parameter (\jep) comes from Eq.\ (\ref{eq:jep}).
The planetary mean density ($\rho\sb{\rm p}$) assumes the observed
mass and transit radius.
$F\sb{\rm XUV}$ (section \ref{sec:hydro}) is the stellar XUV-flux
received by the planet for the given orbital semi-major axis ($a$),
stellar mass ($M\sb{\rm s}$), age, and rotational angular velocity
($\Omega\sb{\rm rot}$).
$L\sb{\rm hy}$ and $L\sb{\rm en}$ (section \ref{sec:hydro}) are the
hydrodynamic and energy-limited XUV-driven ($L\sb{\rm en}$) mass-loss
rates of the selected hydrodynamic-modeled planets (mostly low-density
planets with $\jep < 20$).}
\label{table:allplanets}
\begin{tabular}{@{\extracolsep{\fill}}lr@{}lr@{}lrrrccrrrccrr}
\hline
Name                   & \mctc{$M\sb{\rm p}$}   & \mctc{$R\sb{\rm p}$}   &
$T\sb{\rm eq}$         & \jep          & \multicolumn{1}{c}{$\rho\sb{\rm p}$} &
$a$                    & $M\sb{\rm s}$          & Age                    &
$\Omega\sb{\rm rot}$   & $F\sb{\rm XUV}$        & $L\sb{\rm hy}$         &
$L\sb{\rm en}$         & $L\sb{\rm hy}$/$L\sb{\rm en}$ &
Ref.$\sp{a,b}$ \\
                       & \mctc{$M\sb{\oplus}$}  & \mctc{$R\sb{\oplus}$}  &
K                      &                        & g\,cm\sp{-$3$}         &
AU                     & $M\sb{\odot}$          & Gyr                    &
$\Omega\sb{\odot}$     & erg\,s\sp{-$1$}\,cm\sp{-$2$}  & s$\sp{-1}$      &
s$\sp{-1}$                    &                             \\
\hline
55 Cnc e      &  8.38 & $\pm 0.39$               &  2.08 & $\pm 0.16$             & 1957 &  15.6 &  5.14 & 0.015 & 0.91 &     10.2 &      1.4 & 173333.5 &         $\cdots$ &         $\cdots$ & $\cdots$ & De11$^{\rm R}$\\
CoRoT-22 b    & 12.08 & $\sb{ -8.90}\sp{+13.99}$ &  4.88 & $\sb{-0.39}\sp{+0.17}$ & 1007 &  18.6 &  0.57 & 0.092 & 1.10 &      3.3 &      1.9 &  10959.0 & 3.4\by$\ttt{35}$ & 1.4\by$\ttt{35}$ &      2.3 & Mou14$^{\rm R}$\\
CoRoT-24 c    & 27.97 & $\pm11.12$               &  4.94 & $\pm 0.45$             &  706 &  60.8 &  1.28 & 0.098 & 0.91 &     11.0 &      1.3 &   2429.9 &         $\cdots$ &         $\cdots$ & $\cdots$ & Alo14$^{\rm R}$\\
CoRoT-7 b     &  5.72 & $\pm 0.95$               &  1.58 & $\pm 0.07$             & 1760 &  15.6 &  7.97 & 0.017 & 0.91 &      1.3 &      2.3 & 293392.3 &         $\cdots$ &         $\cdots$ & $\cdots$ & Ba14$^{\rm R}$\\
EPIC 2037 b   & 21.01 & $\pm 5.40$               &  5.69 & $\pm 0.56$             &  776 &  36.1 &  0.63 & 0.154 & 1.12 &      5.0 &      0.4 &    231.6 &         $\cdots$ &         $\cdots$ & $\cdots$ & Pet15$^{\rm R}$\\
EPIC 2037 c   & 27.00 & $\pm 6.90$               &  7.83 & $\pm 0.72$             &  612 &  42.7 &  0.31 & 0.247 & 1.12 &      5.0 &      0.4 &     90.0 &         $\cdots$ &         $\cdots$ & $\cdots$ & Pet15$^{\rm R}$\\
GJ 1132 b     &  1.62 & $\pm 0.54$               &  1.16 & $\pm 0.11$             &  578 &  18.4 &  5.79 & 0.015 & 0.18 &      5.0 &      5.2 & 156736.6 &         $\cdots$ &         $\cdots$ & $\cdots$ & BT15$^{\rm R}$\\
GJ 1214 b     &  6.36 & $\pm 0.95$               &  2.28 & $\pm 0.08$             &  599 &  35.3 &  2.97 & 0.014 & 0.18 &      6.0 &      5.2 & 181777.4 &         $\cdots$ &         $\cdots$ & $\cdots$ & Ha13$^{\rm R}$\\
GJ 3470 b     & 13.67 & $\pm 1.59$               &  3.88 & $\pm 0.33$             &  692 &  38.5 &  1.29 & 0.031 & 0.51 &      2.5 & $\cdots$ &   5353.9 &         $\cdots$ &         $\cdots$ & $\cdots$ & Bi14$^{\rm R}$\\
GJ 436 b      & 22.25 & $\pm 2.22$               &  4.17 & $\pm 0.17$             &  642 &  62.9 &  1.69 & 0.029 & 0.47 &      6.0 &      0.3 &    699.6 & 1.3\by$\ttt{33}$ & 8.9\by$\ttt{33}$ & 0.1 & Mac14$^{\rm R}$\\
HAT-P-11 b    & 25.75 & $\pm 2.86$               &  4.74 & $\pm 0.16$             &  866 &  47.6 &  1.34 & 0.053 & 0.81 &      6.5 &      1.1 &   5944.9 & 2.7\by$\ttt{34}$ & 2.4\by$\ttt{34}$ &      1.1 & Bak10$^{\rm R}$\\
HAT-P-26 b    & 18.75 & $\pm 2.22$               &  6.34 & $\sb{-0.36}\sp{+0.81}$ &  994 &  22.6 &  0.41 & 0.048 & 0.82 &      9.0 &      1.2 &  12237.8 & 4.4\by$\ttt{35}$ & 2.4\by$\ttt{35}$ &      1.8 & Ha11$^{\rm R}$\\
HD 219134 b   &  4.46 & $\pm 0.47$               &  1.61 & $\pm 0.09$             & 1022 &  20.6 &  5.94 & 0.038 & 0.79 &     12.5 &      0.3 &    926.6 &         $\cdots$ &         $\cdots$ & $\cdots$ & Mot15$^{\rm R}$\\
HD 97658 b    &  7.63 & $\sb{ -0.64}\sp{ +0.95}$ &  2.24 & $\sb{-0.09}\sp{+0.10}$ &  758 &  34.0 &  3.72 & 0.080 & 0.77 &      6.0 &      0.4 &    485.1 &         $\cdots$ &         $\cdots$ & $\cdots$ & VG14$^{\rm R}$\\
HIP 116454 b  & 11.76 & $\pm 1.27$               &  2.54 & $\pm 0.18$             &  691 &  50.8 &  3.98 & 0.091 & 0.78 &      2.0 &      1.5 &   4879.7 &         $\cdots$ &         $\cdots$ & $\cdots$ & V15$^{\rm R}$\\
K2-19 c       & 15.90 & $\sb{ -2.80}\sp{ +7.70}$ &  4.51 & $\pm 0.47$             &  783 &  34.1 &  0.96 & 0.100 & 0.95 & $\cdots$ &      0.5 &    540.3 & 6.4\by$\ttt{33}$ & 3.2\by$\ttt{33}$ &      2.0 & Ba15$^{\rm R,T}$\\
K2-3 b        &  8.40 & $\pm 2.10$               &  2.08 & $\sb{-0.09}\sp{+0.18}$ &  500 &  61.3 &  5.18 & 0.077 & 0.61 &      2.0 & $\cdots$ &   1232.2 &         $\cdots$ &         $\cdots$ & $\cdots$ & Alm15$^{\rm R}$\\
K2-3 c        &  2.10 & $\sb{ -1.30}\sp{ +2.10}$ &  1.65 & $\sb{-0.07}\sp{+0.16}$ &  371 &  26.0 &  2.58 & 0.141 & 0.61 &      2.0 & $\cdots$ &    374.9 &         $\cdots$ &         $\cdots$ & $\cdots$ & Alm15$^{\rm R}$\\
K2-3 d        & 11.10 & $\pm 3.50$               &  1.53 & $\pm 0.11$             &  305 & 180.8 & 17.22 & 0.209 & 0.61 &      2.0 & $\cdots$ &    170.1 &         $\cdots$ &         $\cdots$ & $\cdots$ & Alm15$^{\rm R}$\\
K2-56 b     & 16.30 & $\sb{ -6.10}\sp{ +6.00}$ &  2.23 & $\sb{-0.11}\sp{+0.14}$ &  545 & 101.6 &  8.11 & 0.241 & 0.96 &      3.3 &      1.9 &   1517.7 &         $\cdots$ &         $\cdots$ & $\cdots$ & E16$^{\rm R}$\\
Kepler-10 b   &  3.72 & $\pm 0.42$               &  1.48 & $\sb{-0.03}\sp{+0.04}$ & 2129 &   8.9 &  6.31 & 0.017 & 0.91 &      9.1 &      1.0 & 124224.8 &         $\cdots$ &         $\cdots$ & $\cdots$ & W16$^{\rm R}$\\
Kepler-10 c   & 13.98 & $\pm 1.79$               &  2.36 & $\sb{-0.04}\sp{+0.09}$ &  570 &  78.8 &  5.89 & 0.240 & 0.91 &      9.1 &      1.0 &    639.6 &         $\cdots$ &         $\cdots$ & $\cdots$ & W16$^{\rm R}$\\
Kepler-100 b  &  7.31 & $\pm 3.18$               &  1.32 & $\pm 0.04$             & 1271 &  32.9 & 17.37 & 0.073 & 1.08 &      6.9 &      1.3 &  14282.2 &         $\cdots$ &         $\cdots$ & $\cdots$ & Mar14$^{\rm R}$\\
Kepler-102 d  &  3.81 & $\pm 1.91$               &  1.18 & $\pm 0.04$             &  701 &  35.0 & 12.86 & 0.086 & 0.81 &      4.1 &      0.4 &    302.8 &         $\cdots$ &         $\cdots$ & $\cdots$ & Mar14$^{\rm R}$\\
Kepler-102 e  &  8.90 & $\pm 1.91$               &  2.22 & $\pm 0.07$             &  604 &  50.3 &  4.47 & 0.117 & 0.81 &      4.1 &      0.4 &    166.5 &         $\cdots$ &         $\cdots$ & $\cdots$ & Mar14$^{\rm R}$\\
Kepler-103 b  &  9.85 & $\pm 8.58$               &  3.38 & $\pm 0.09$             &  946 &  23.4 &  1.41 & 0.128 & 1.09 &      6.0 &      0.9 &   2165.8 & 3.1\by$\ttt{34}$ & 9.0\by$\ttt{33}$ &      3.4 & Mar14$^{\rm R}$\\
Kepler-104 b  & 19.60 & $\sb{-12.40}\sp{+14.50}$ &  3.13 & $\sb{-0.70}\sp{+0.52}$ & 1043 &  45.5 &  3.53 & 0.094 & 0.81 &      3.7 & $\cdots$ &   5720.3 &         $\cdots$ &         $\cdots$ & $\cdots$ & HL14$^{\rm T}$\\
Kepler-105 b  &  3.70 & $\pm 2.00$               &  2.22 & $\pm 0.11$             & 1088 &  11.6 &  1.87 & 0.060 & 0.96 &      3.5 &      1.6 &  17615.2 & 3.8\by$\ttt{35}$ & 2.1\by$\ttt{35}$ &      1.8 & JH15b$^{\rm T}$\\
Kepler-105 c  &  4.60 & $\pm 0.90$               &  1.31 & $\pm 0.07$             &  993 &  26.8 & 11.28 & 0.072 & 0.96 &      3.5 &      1.6 &  12198.6 &         $\cdots$ &         $\cdots$ & $\cdots$ & JH15b$^{\rm T}$\\
Kepler-106 c  & 10.49 & $\pm 3.18$               &  2.50 & $\pm 0.33$             &  863 &  36.8 &  3.69 & 0.111 & 1.00 &      3.3 &      0.2 &     68.5 &         $\cdots$ &         $\cdots$ & $\cdots$ & Mar14$^{\rm R}$\\
Kepler-106 e  & 11.12 & $\pm 5.72$               &  2.56 & $\pm 0.33$             &  583 &  56.5 &  3.66 & 0.243 & 1.00 &      3.3 &      0.2 &     14.3 &         $\cdots$ &         $\cdots$ & $\cdots$ & Mar14$^{\rm R}$\\
Kepler-11 b   &  1.91 & $\sb{ -0.95}\sp{ +1.27}$ &  1.81 & $\sb{-0.04}\sp{+0.03}$ &  931 &   8.6 &  1.78 & 0.091 & 0.96 &      6.9 &      0.2 &    175.8 & 2.2\by$\ttt{35}$ & 8.7\by$\ttt{33}$ &     25.3 & L13$^{\rm T}$\\
Kepler-11 c   &  2.86 & $\sb{ -1.59}\sp{ +2.86}$ &  2.87 & $\sb{-0.06}\sp{+0.04}$ &  859 &   8.8 &  0.67 & 0.107 & 0.96 &      6.9 &      0.2 &    127.2 & 4.0\by$\ttt{35}$ & 1.5\by$\ttt{34}$ &     26.7 & L13$^{\rm T}$\\
Kepler-11 d   &  7.31 & $\sb{ -1.59}\sp{ +0.95}$ &  3.12 & $\sb{-0.07}\sp{+0.06}$ &  714 &  24.9 &  1.33 & 0.155 & 0.96 &      6.9 &      0.2 &     60.6 & 3.6\by$\ttt{32}$ & 3.5\by$\ttt{32}$ &      1.0 & L13$^{\rm T}$\\
Kepler-11 e   &  7.95 & $\sb{ -2.22}\sp{ +1.59}$ &  4.20 & $\sb{-0.09}\sp{+0.07}$ &  636 &  22.6 &  0.59 & 0.195 & 0.96 &      6.9 &      0.2 &     38.3 & 2.0\by$\ttt{33}$ & 4.9\by$\ttt{32}$ &      4.1 & L13$^{\rm T}$\\
Kepler-11 f   &  1.91 & $\pm 0.95$               &  2.49 & $\sb{-0.07}\sp{+0.04}$ &  562 &  10.3 &  0.68 & 0.250 & 0.96 &      6.9 &      0.2 &     23.3 & 8.4\by$\ttt{34}$ & 2.8\by$\ttt{33}$ &     30.0 & L13$^{\rm T}$\\
Kepler-113 b  & 11.76 & $\pm 4.13$               &  1.82 & $\pm 0.04$             &  844 &  58.1 & 10.80 & 0.050 & 0.75 &      3.2 &      0.3 &    635.2 &         $\cdots$ &         $\cdots$ & $\cdots$ & Mar14$^{\rm R}$\\
Kepler-114 c  &  2.86 & $\pm 0.64$               &  1.60 & $\pm 0.18$             &  714 &  18.9 &  3.82 & 0.065 & 0.56 &      2.7 &      0.9 &   4657.9 &         $\cdots$ &         $\cdots$ & $\cdots$ & X14$^{\rm T}$\\
Kepler-114 d  &  3.81 & $\pm 1.59$               &  2.54 & $\pm 0.28$             &  628 &  18.1 &  1.29 & 0.083 & 0.56 &      2.7 &      0.9 &   2796.6 & 6.1\by$\ttt{34}$ & 1.8\by$\ttt{34}$ &      3.4 & X14$^{\rm T}$\\
Kepler-120 b  &  8.50 & $\sb{ -7.50}\sp{ +9.70}$ &  2.31 & $\pm 0.37$             &  613 &  45.5 &  3.80 & 0.055 & 0.59 &      3.6 & $\cdots$ &   1524.5 &         $\cdots$ &         $\cdots$ & $\cdots$ & HL14$^{\rm T}$\\
Kepler-122 e  & 27.70 & $\sb{ -9.80}\sp{+11.40}$ &  2.02 & $\sb{-0.19}\sp{+0.70}$ &  676 & 153.8 & 18.53 & 0.227 & 0.99 &      3.9 & $\cdots$ &   1104.9 &         $\cdots$ &         $\cdots$ & $\cdots$ & HL14$^{\rm T}$\\
Kepler-131 b  & 16.21 & $\pm 3.50$               &  2.41 & $\pm 0.20$             &  785 &  64.9 &  6.37 & 0.126 & 1.02 &      3.7 &      0.2 &     79.2 &         $\cdots$ &         $\cdots$ & $\cdots$ & Mar14$^{\rm R}$\\
Kepler-131 c  &  8.26 & $\pm 6.04$               &  0.84 & $\pm 0.07$             &  673 & 110.6 & 76.46 & 0.171 & 1.02 &      3.7 &      0.2 &     42.9 &         $\cdots$ &         $\cdots$ & $\cdots$ & Mar14$^{\rm R}$\\
Kepler-136 b  & 19.80 & $\sb{-10.40}\sp{+11.70}$ &  1.80 & $\sb{-0.17}\sp{+0.35}$ & 1060 &  78.6 & 18.72 & 0.106 & 1.20 &      2.8 & $\cdots$ &  12210.7 &         $\cdots$ &         $\cdots$ & $\cdots$ & HL14$^{\rm T}$\\
Kepler-138 b  &  0.07 & $\sb{ -0.04}\sp{ +0.06}$ &  0.53 & $\pm 0.03$             &  449 &   2.1 &  2.51 & 0.075 & 0.52 &      4.7 &      1.4 &   2705.8 &         $\cdots$ &         $\cdots$ & $\cdots$ & JH15a$^{\rm T}$\\
Kepler-138 c  &  1.97 & $\sb{ -1.12}\sp{ +1.91}$ &  1.20 & $\pm 0.07$             &  408 &  30.5 &  6.28 & 0.090 & 0.52 &      4.7 &      1.4 &   1838.5 &         $\cdots$ &         $\cdots$ & $\cdots$ & JH15a$^{\rm T}$\\
Kepler-138 d  &  0.64 & $\sb{ -0.39}\sp{ +0.67}$ &  1.21 & $\pm 0.08$             &  343 &  11.6 &  1.98 & 0.128 & 0.52 &      4.7 &      1.4 &    924.8 & 8.4\by$\ttt{34}$ & 1.2\by$\ttt{34}$ &      7.1 & JH15a$^{\rm T}$\\
\hline
\end{tabular}
\end{table*}

\begin{table*}
\contcaption{Sub-Neptune Planet Parameters.}
\begin{tabular}{@{\extracolsep{\fill}}lr@{}lr@{}lrrrccrrrccrr}
\hline
Name                   & \mctc{$M\sb{\rm p}$}   & \mctc{$R\sb{\rm p}$}   &
$T\sb{\rm eq}$         & \jep          & \multicolumn{1}{c}{$\rho\sb{\rm p}$} &
$a$                    & $M\sb{\rm s}$          & Age                    &
$\Omega\sb{\rm rot}$   & $F\sb{\rm XUV}$        & $L\sb{\rm hy}$         &
$L\sb{\rm en}$         & $L\sb{\rm hy}$/$L\sb{\rm en}$ &
Ref.$\sp{a,b}$ \\
                       & \mctc{$M\sb{\oplus}$}  & \mctc{$R\sb{\oplus}$}  &
K                      &                        & g\,cm$\sp{-3}$         &
AU                     & $M\sb{\odot}$          & Gyr                    &
$\Omega\sb{\odot}$     & erg\,s\sp{-$1$}\,cm\sp{-$2$}  & s$\sp{-1}$      &
s$\sp{-1}$                    &                             \\
\hline
Kepler-161 b  & 12.10 & $\sb{ -6.30}\sp{ +7.40}$ &  1.93 & $\sb{-0.14}\sp{+0.31}$ &  948 &  50.1 &  9.28 & 0.054 & 0.77 &      4.7 &      1.0 &   7681.3 &         $\cdots$ &         $\cdots$ & $\cdots$ & HL14$^{\rm T}$\\
Kepler-161 c  & 11.80 & $\sb{ -7.50}\sp{+10.50}$ &  1.87 & $\sb{-0.14}\sp{+0.30}$ &  844 &  56.6 &  9.95 & 0.068 & 0.77 &      4.7 &      1.0 &   4844.0 &         $\cdots$ &         $\cdots$ & $\cdots$ & HL14$^{\rm T}$\\
Kepler-176 c  & 23.00 & $\sb{ -8.00}\sp{+13.50}$ &  2.56 & $\sb{-0.26}\sp{+0.93}$ &  745 &  91.4 &  7.56 & 0.102 & 0.83 &      4.7 &      0.8 &   1567.5 &         $\cdots$ &         $\cdots$ & $\cdots$ & HL14$^{\rm T}$\\
Kepler-176 d  & 15.20 & $\sb{ -5.80}\sp{+10.40}$ &  2.47 & $\sb{-0.25}\sp{+0.90}$ &  589 &  79.2 &  5.56 & 0.163 & 0.83 &      4.7 &      0.8 &    613.8 &         $\cdots$ &         $\cdots$ & $\cdots$ & HL14$^{\rm T}$\\
Kepler-177 b  &  1.91 & $\sb{ -0.32}\sp{ +0.64}$ &  2.91 & $\sb{-0.30}\sp{+1.53}$ &  655 &   7.6 &  0.43 & 0.222 & 1.07 &      4.4 & $\cdots$ &    962.8 & 4.1\by$\ttt{36}$ & 2.0\by$\ttt{35}$ &     20.5 & X14$^{\rm T}$\\
Kepler-177 c  &  7.63 & $\sb{ -3.18}\sp{ +3.50}$ &  7.10 & $\sb{-0.72}\sp{+3.71}$ &  594 &  13.7 &  0.12 & 0.270 & 1.07 &      4.4 & $\cdots$ &    651.5 & 5.8\by$\ttt{35}$ & 1.0\by$\ttt{35}$ &      5.8 & X14$^{\rm T}$\\
Kepler-18 b   &  6.99 & $\pm 3.50$               &  2.00 & $\pm 0.10$             & 1284 &  20.7 &  4.84 & 0.045 & 0.97 &     10.7 &      1.9 &  46258.8 &         $\cdots$ &         $\cdots$ & $\cdots$ & Co11$^{\rm R,T}$\\
Kepler-18 c   & 17.16 & $\pm 1.91$               &  5.50 & $\pm 0.26$             &  990 &  23.9 &  0.57 & 0.075 & 0.97 &     10.7 &      1.9 &  16344.6 & 3.3\by$\ttt{35}$ & 2.1\by$\ttt{35}$ &      1.6 & Co11$^{\rm R,T}$\\
Kepler-18 d   & 16.53 & $\pm 1.27$               &  6.99 & $\pm 0.33$             &  793 &  22.6 &  0.27 & 0.117 & 0.97 &     10.7 &      1.9 &   6729.1 & 3.5\by$\ttt{35}$ & 1.9\by$\ttt{35}$ &      1.8 & Co11$^{\rm R,T}$\\
Kepler-184 c  &  8.80 & $\sb{ -5.70}\sp{ +7.40}$ &  1.95 & $\sb{-0.19}\sp{+0.84}$ &  693 &  49.4 &  6.55 & 0.141 & 0.87 &      4.5 & $\cdots$ &   1017.7 &         $\cdots$ &         $\cdots$ & $\cdots$ & HL14$^{\rm T}$\\
Kepler-189 c  & 22.70 & $\sb{-10.60}\sp{+17.10}$ &  2.40 & $\sb{-0.18}\sp{+0.52}$ &  590 & 121.5 &  9.06 & 0.137 & 0.84 &      5.2 & $\cdots$ &    497.7 &         $\cdots$ &         $\cdots$ & $\cdots$ & HL14$^{\rm T}$\\
Kepler-197 c  &  5.30 & $\sb{ -2.90}\sp{ +3.30}$ &  1.23 & $\pm 0.04$             & 1021 &  32.0 & 15.71 & 0.090 & 0.92 &      5.4 & $\cdots$ &   3777.0 &         $\cdots$ &         $\cdots$ & $\cdots$ & HL14$^{\rm T}$\\
Kepler-20 b   &  8.58 & $\pm 2.22$               &  1.91 & $\sb{-0.21}\sp{+0.12}$ & 1197 &  28.5 &  6.82 & 0.045 & 0.91 &      5.2 &      0.2 &    732.8 &         $\cdots$ &         $\cdots$ & $\cdots$ & Ga12$^{\rm R}$\\
Kepler-20 c   & 16.21 & $\sb{ -3.81}\sp{ +3.18}$ &  3.07 & $\sb{-0.31}\sp{+0.20}$ &  836 &  47.8 &  3.08 & 0.093 & 0.91 &      5.2 &      0.2 &    174.6 &         $\cdots$ &         $\cdots$ & $\cdots$ & Ga12$^{\rm R}$\\
Kepler-211 c  & 18.30 & $\sb{-17.00}\sp{+22.40}$ &  2.45 & $\sb{-1.09}\sp{+1.62}$ &  898 &  63.1 &  6.86 & 0.062 & 0.97 &      1.6 &      2.0 &  12588.9 &         $\cdots$ &         $\cdots$ & $\cdots$ & HL14$^{\rm T}$\\
Kepler-215 d  & 23.60 & $\sb{-11.90}\sp{+17.30}$ &  2.34 & $\sb{-0.34}\sp{+0.44}$ &  652 & 117.1 & 10.16 & 0.185 & 0.77 &      1.6 & $\cdots$ &   2352.6 &         $\cdots$ &         $\cdots$ & $\cdots$ & HL14$^{\rm T}$\\
Kepler-219 d  & 19.10 & $\sb{-17.70}\sp{+29.90}$ &  2.51 & $\sb{-0.40}\sp{+0.63}$ &  652 &  88.4 &  6.66 & 0.272 & 1.15 &      4.6 &      1.6 &   1158.9 &         $\cdots$ &         $\cdots$ & $\cdots$ & HL14$^{\rm T}$\\
Kepler-221 c  & 15.10 & $\sb{ -6.70}\sp{+10.10}$ &  2.68 & $\sb{-0.26}\sp{+0.61}$ &  942 &  45.3 &  4.33 & 0.059 & 0.72 &      4.6 &      2.9 &  70313.6 &         $\cdots$ &         $\cdots$ & $\cdots$ & HL14$^{\rm T}$\\
Kepler-226 b  & 24.00 & $\sb{-10.10}\sp{+11.80}$ &  1.56 & $\sb{-0.12}\sp{+0.58}$ & 1108 & 105.3 & 34.86 & 0.047 & 0.86 &      4.4 & $\cdots$ &   7076.1 &         $\cdots$ &         $\cdots$ & $\cdots$ & HL14$^{\rm T}$\\
Kepler-23 b   & 15.20 & $\sb{ -2.90}\sp{ +3.20}$ &  1.90 & $\pm 0.06$             & 1277 &  47.5 & 12.22 & 0.075 & 1.11 &      6.6 & $\cdots$ &   8385.5 &         $\cdots$ &         $\cdots$ & $\cdots$ & HL14$^{\rm T}$\\
Kepler-23 d   & 17.60 & $\sb{-11.90}\sp{+13.70}$ &  2.20 & $\pm 0.07$             &  993 &  61.1 &  9.12 & 0.124 & 1.11 &      6.6 & $\cdots$ &   3067.7 &         $\cdots$ &         $\cdots$ & $\cdots$ & HL14$^{\rm T}$\\
Kepler-231 b  &  4.90 & $\sb{ -1.30}\sp{ +1.80}$ &  1.82 & $\sb{-0.25}\sp{+0.26}$ &  462 &  44.1 &  4.48 & 0.074 & 0.51 &      3.5 &      1.2 &   2185.7 &         $\cdots$ &         $\cdots$ & $\cdots$ & K14$^{\rm T}$\\
Kepler-231 c  &  2.20 & $\sb{ -1.10}\sp{ +1.50}$ &  1.69 & $\sb{-0.23}\sp{+0.24}$ &  372 &  26.5 &  2.51 & 0.114 & 0.51 &      3.5 &      1.2 &    921.0 &         $\cdots$ &         $\cdots$ & $\cdots$ & K14$^{\rm T}$\\
Kepler-238 f  & 13.35 & $\sb{ -2.54}\sp{ +2.86}$ &  2.00 & $\sb{-0.17}\sp{+0.85}$ &  635 &  79.8 &  9.24 & 0.272 & 1.06 &      6.8 & $\cdots$ &    507.0 &         $\cdots$ &         $\cdots$ & $\cdots$ & X14$^{\rm T}$\\
Kepler-244 d  & 15.20 & $\sb{-13.00}\sp{+20.00}$ &  2.32 & $\sb{-0.18}\sp{+0.83}$ &  640 &  77.6 &  6.71 & 0.140 & 0.88 &      4.8 & $\cdots$ &    720.1 &         $\cdots$ &         $\cdots$ & $\cdots$ & HL14$^{\rm T}$\\
Kepler-245 d  & 21.60 & $\sb{-12.10}\sp{+16.60}$ &  3.31 & $\sb{-0.37}\sp{+1.40}$ &  489 & 101.1 &  3.28 & 0.202 & 0.80 &      3.6 &      1.4 &    987.3 &         $\cdots$ &         $\cdots$ & $\cdots$ & HL14$^{\rm T}$\\
Kepler-25 b   &  9.60 & $\pm 4.20$               &  2.71 & $\pm 0.05$             & 1305 &  20.6 &  2.65 & 0.070 & 1.19 &      2.9 &      3.9 &  92614.7 &         $\cdots$ &         $\cdots$ & $\cdots$ & Mar14$^{\rm R}$\\
Kepler-25 c   & 24.60 & $\pm 5.70$               &  5.21 & $\pm 0.09$             & 1029 &  34.8 &  0.96 & 0.113 & 1.19 &      2.9 &      3.9 &  35845.4 & 1.6\by$\ttt{35}$ & 1.7\by$\ttt{35}$ &      0.9 & Mar14$^{\rm R}$\\
Kepler-254 c  &  3.20 & $\sb{ -2.70}\sp{ +3.70}$ &  2.09 & $\sb{-0.16}\sp{+0.84}$ &  845 &  13.7 &  1.93 & 0.105 & 0.97 &      4.0 &      1.1 &   2984.8 & 9.1\by$\ttt{34}$ & 2.9\by$\ttt{34}$ &      3.1 & HL14$^{\rm T}$\\
Kepler-26 b   &  5.10 & $\pm 0.70$               &  2.78 & $\pm 0.11$             &  465 &  29.9 &  1.31 & 0.085 & 0.54 &      3.0 &      2.0 &   5260.6 & 2.7\by$\ttt{34}$ & 2.8\by$\ttt{34}$ &      1.0 & JH15b$^{\rm T}$\\
Kepler-26 c   &  6.20 & $\pm 0.70$               &  2.72 & $\pm 0.12$             &  415 &  41.6 &  1.70 & 0.107 & 0.54 &      3.0 &      2.0 &   3338.5 & 5.8\by$\ttt{33}$ & 1.7\by$\ttt{34}$ &      0.3 & JH15b$^{\rm T}$\\
Kepler-27 c   & 21.20 & $\sb{ -3.70}\sp{ +3.20}$ &  7.17 & $\sb{-0.27}\sp{+0.38}$ &  457 &  49.0 &  0.32 & 0.191 & 0.65 &      1.6 &      0.5 &    222.8 &         $\cdots$ &         $\cdots$ & $\cdots$ & HL14$^{\rm T}$\\
Kepler-276 c  & 16.53 & $\sb{ -3.50}\sp{ +4.45}$ &  2.91 & $\sb{-0.28}\sp{+1.27}$ &  669 &  64.4 &  3.71 & 0.203 & 1.10 &      3.8 & $\cdots$ &   1237.2 &         $\cdots$ &         $\cdots$ & $\cdots$ & X14$^{\rm T}$\\
Kepler-276 d  & 16.21 & $\sb{ -4.45}\sp{ +5.09}$ &  2.81 & $\sb{-0.27}\sp{+1.23}$ &  581 &  75.4 &  4.05 & 0.269 & 1.10 &      3.8 & $\cdots$ &    704.2 &         $\cdots$ &         $\cdots$ & $\cdots$ & X14$^{\rm T}$\\
Kepler-28 b   &  8.80 & $\sb{ -3.10}\sp{ +3.80}$ &  2.93 & $\pm 0.46$             &  743 &  30.6 &  1.93 & 0.062 & 0.75 &      2.2 &      1.5 &   7830.8 &         $\cdots$ &         $\cdots$ & $\cdots$ & HL14$^{\rm T}$\\
Kepler-28 c   & 10.90 & $\sb{ -4.50}\sp{ +6.10}$ &  2.77 & $\pm 0.44$             &  650 &  45.9 &  2.83 & 0.081 & 0.75 &      2.2 &      1.5 &   4587.9 &         $\cdots$ &         $\cdots$ & $\cdots$ & HL14$^{\rm T}$\\
Kepler-289 b  &  7.31 & $\pm 6.67$               &  2.15 & $\pm 0.10$             &  630 &  40.8 &  4.03 & 0.210 & 1.08 &      3.8 &      3.0 &   4490.4 &         $\cdots$ &         $\cdots$ & $\cdots$ & S14$^{\rm T}$\\
Kepler-289 d  &  4.13 & $\pm 0.95$               &  2.68 & $\pm 0.17$             &  502 &  23.2 &  1.18 & 0.330 & 1.08 &      3.8 &      3.0 &   1818.4 & 4.1\by$\ttt{34}$ & 1.9\by$\ttt{34}$ &      2.2 & S14$^{\rm T}$\\
Kepler-29 b   &  4.50 & $\pm 1.50$               &  3.35 & $\pm 0.22$             &  872 &  11.7 &  0.66 & 0.092 & 0.98 & $\cdots$ &      2.3 &  13666.1 & 1.3\by$\ttt{36}$ & 4.0\by$\ttt{35}$ &      3.3 & JH15b$^{\rm T}$\\
Kepler-29 c   &  4.00 & $\pm 1.30$               &  3.14 & $\pm 0.20$             &  802 &  12.0 &  0.71 & 0.109 & 0.98 & $\cdots$ &      2.3 &   9778.1 & 8.9\by$\ttt{35}$ & 2.8\by$\ttt{35}$ &      3.2 & JH15b$^{\rm T}$\\
Kepler-30 b   & 11.30 & $\pm 1.40$               &  3.90 & $\pm 0.20$             &  599 &  36.7 &  1.05 & 0.186 & 0.99 &      1.6 &      1.1 &    718.8 & 7.5\by$\ttt{33}$ & 4.3\by$\ttt{33}$ &      1.7 & SO12$^{\rm T}$\\
Kepler-30 d   & 23.10 & $\pm 2.70$               &  8.79 & $\pm 0.50$             &  353 &  56.4 &  0.19 & 0.534 & 0.99 &      1.6 &      1.1 &     86.7 &         $\cdots$ &         $\cdots$ & $\cdots$ & SO12$^{\rm T}$\\
Kepler-305 b  & 10.49 & $\sb{ -1.91}\sp{ +2.54}$ &  3.60 & $\sb{-0.36}\sp{+0.90}$ &  927 &  23.8 &  1.24 & 0.056 & 0.76 &      4.4 & $\cdots$ &   3742.0 & 5.1\by$\ttt{34}$ & 2.4\by$\ttt{34}$ &      2.1 & X14$^{\rm T}$\\
Kepler-305 c  &  6.04 & $\sb{ -2.22}\sp{ +2.54}$ &  3.30 & $\sb{-0.33}\sp{+0.82}$ &  808 &  17.2 &  0.93 & 0.073 & 0.76 &      4.4 & $\cdots$ &   2158.9 & 8.0\by$\ttt{34}$ & 2.5\by$\ttt{34}$ &      3.2 & X14$^{\rm T}$\\
Kepler-307 b  &  3.18 & $\pm 0.64$               &  3.20 & $\sb{-0.46}\sp{+1.20}$ &  883 &   8.5 &  0.54 & 0.093 & 0.98 &      6.6 &      2.0 &  11054.6 & 3.9\by$\ttt{36}$ & 9.5\by$\ttt{35}$ &      4.1 & X14$^{\rm T}$\\
Kepler-307 c  &  1.59 & $\sb{ -0.32}\sp{ +0.64}$ &  2.81 & $\sb{-0.42}\sp{+1.05}$ &  818 &   5.2 &  0.40 & 0.108 & 0.98 &      6.6 &      2.0 &   8159.4 & 9.8\by$\ttt{37}$ & 1.2\by$\ttt{36}$ &     81.7 & X14$^{\rm T}$\\
Kepler-31 c   & 29.50 & $\sb{ -7.00}\sp{ +9.60}$ &  4.71 & $\sb{-0.57}\sp{+2.23}$ &  653 &  72.7 &  1.56 & 0.260 & 1.21 &      4.2 & $\cdots$ &   1108.1 &         $\cdots$ &         $\cdots$ & $\cdots$ & HL14$^{\rm T}$\\
Kepler-32 b   &  9.40 & $\sb{ -3.10}\sp{ +3.60}$ &  2.25 & $\pm 0.11$             &  595 &  53.2 &  4.55 & 0.050 & 0.54 &      2.7 &      0.8 &   2232.3 &         $\cdots$ &         $\cdots$ & $\cdots$ & HL14$^{\rm T}$\\
Kepler-32 c   &  7.70 & $\sb{ -3.80}\sp{ +5.00}$ &  2.02 & $\pm 0.11$             &  443 &  65.1 &  5.15 & 0.090 & 0.54 &      2.7 &      0.8 &    689.0 &         $\cdots$ &         $\cdots$ & $\cdots$ & HL14$^{\rm T}$\\
Kepler-326 c  & 17.40 & $\sb{-10.70}\sp{+15.40}$ &  2.79 & $\sb{-1.31}\sp{+1.96}$ &  974 &  48.5 &  4.42 & 0.051 & 0.98 &      4.7 &      2.8 &  34240.8 &         $\cdots$ &         $\cdots$ & $\cdots$ & HL14$^{\rm T}$\\
Kepler-326 d  &  6.90 & $\sb{ -5.90}\sp{ +8.50}$ &  2.41 & $\sb{-1.13}\sp{+1.70}$ &  856 &  25.3 &  2.72 & 0.066 & 0.98 &      4.7 &      2.8 &  20445.5 & 4.7\by$\ttt{34}$ & 6.1\by$\ttt{34}$ &      0.8 & HL14$^{\rm T}$\\
\hline
\end{tabular}
\end{table*}

\begin{table*}
\contcaption{Sub-Neptune Planet Parameters.}
\begin{tabular}{@{\extracolsep{\fill}}lr@{}lr@{}lrrrccrrrccrr}
\hline
Name                   & \mctc{$M\sb{\rm p}$}   & \mctc{$R\sb{\rm p}$}   &
$T\sb{\rm eq}$         & \jep          & \multicolumn{1}{c}{$\rho\sb{\rm p}$} &
$a$                    & $M\sb{\rm s}$          & Age                    &
$\Omega\sb{\rm rot}$   & $F\sb{\rm XUV}$        & $L\sb{\rm hy}$         &
$L\sb{\rm en}$         & $L\sb{\rm hy}$/$L\sb{\rm en}$ &
Ref.$\sp{a,b}$ \\
                       & \mctc{$M\sb{\oplus}$}  & \mctc{$R\sb{\oplus}$}  &
K                      &                        & g\,cm$\sp{-3}$         &
AU                     & $M\sb{\odot}$          & Gyr                    &
$\Omega\sb{\odot}$     & erg\,s\sp{-$1$}\,cm\sp{-$2$}  & s$\sp{-1}$      &
s$\sp{-1}$                    &                             \\
\hline
Kepler-327 c  & 20.30 & $\sb{-17.70}\sp{+27.10}$ &  1.18 & $\pm 0.11$             &  591 & 220.6 & 68.14 & 0.047 & 0.55 &      3.0 &      1.1 &   4264.3 &         $\cdots$ &         $\cdots$ & $\cdots$ & HL14$^{\rm T}$\\
Kepler-328 b  & 28.61 & $\sb{-12.40}\sp{+13.03}$ &  2.30 & $\sb{-0.24}\sp{+0.97}$ &  627 & 150.3 & 12.96 & 0.219 & 1.15 &      2.6 & $\cdots$ &   1487.0 &         $\cdots$ &         $\cdots$ & $\cdots$ & X14$^{\rm T}$\\
Kepler-33 c   &  0.80 & $\sb{ -0.70}\sp{ +2.50}$ &  3.20 & $\pm 0.30$             & 1113 &   1.7 &  0.13 & 0.119 & 1.29 &      4.4 &      1.0 &   2707.6 & 2.2\by$\ttt{42}$ & 1.4\by$\ttt{36}$ & 1.6\by$\ttt{6}$ & HL16$^{\rm T}$\\
Kepler-33 d   &  4.70 & $\pm 2.00$               &  5.35 & $\pm 0.49$             &  941 &   7.1 &  0.17 & 0.166 & 1.29 &      4.4 &      1.0 &   1385.8 & 1.8\by$\ttt{37}$ & 7.5\by$\ttt{35}$ &     24.0 & HL16$^{\rm T}$\\
Kepler-33 e   &  6.70 & $\sb{ -1.30}\sp{ +1.20}$ &  4.03 & $\pm 0.38$             &  830 &  15.2 &  0.57 & 0.214 & 1.29 &      4.4 &      1.0 &    837.4 & 3.3\by$\ttt{34}$ & 2.8\by$\ttt{34}$ &      1.2 & HL16$^{\rm T}$\\
Kepler-33 f   & 11.50 & $\sb{ -2.10}\sp{ +1.80}$ &  4.47 & $\pm 0.42$             &  762 &  25.6 &  0.71 & 0.254 & 1.29 &      4.4 &      1.0 &    595.6 & 1.8\by$\ttt{34}$ & 6.1\by$\ttt{33}$ &      3.0 & HL16$^{\rm T}$\\
Kepler-333 b  & 28.20 & $\sb{-24.10}\sp{+29.00}$ &  1.61 & $\pm 0.17$             &  506 & 262.1 & 37.27 & 0.087 & 0.54 &      3.2 &      1.1 &   2381.1 &         $\cdots$ &         $\cdots$ & $\cdots$ & HL14$^{\rm T}$\\
Kepler-338 e  &  8.50 & $\sb{ -6.30}\sp{ +7.20}$ &  1.56 & $\pm 0.07$             & 1258 &  32.8 & 12.35 & 0.090 & 1.10 &      4.8 & $\cdots$ &  10799.4 &         $\cdots$ &         $\cdots$ & $\cdots$ & HL14$^{\rm T}$\\
Kepler-339 c  &  7.30 & $\sb{ -6.20}\sp{ +7.80}$ &  1.16 & $\sb{-0.09}\sp{+0.45}$ &  924 &  51.6 & 25.79 & 0.069 & 0.84 &      4.4 &      1.6 &  13691.2 &         $\cdots$ &         $\cdots$ & $\cdots$ & HL14$^{\rm T}$\\
Kepler-339 d  & 14.70 & $\sb{-10.00}\sp{+14.10}$ &  1.18 & $\sb{-0.09}\sp{+0.46}$ &  804 & 117.4 & 49.34 & 0.091 & 0.84 &      4.4 &      1.6 &   7871.5 &         $\cdots$ &         $\cdots$ & $\cdots$ & HL14$^{\rm T}$\\
Kepler-350 c  &  6.04 & $\pm 3.18$               &  3.11 & $\sb{-0.61}\sp{+1.43}$ & 1009 &  14.6 &  1.11 & 0.134 & 1.00 &      3.2 &      8.5 & 243252.4 & 1.6\by$\ttt{36}$ & 1.8\by$\ttt{36}$ &      0.9 & X14$^{\rm T}$\\
Kepler-350 d  & 14.94 & $\sb{ -4.77}\sp{ +5.40}$ &  2.81 & $\sb{-0.54}\sp{+1.28}$ &  888 &  45.4 &  3.73 & 0.172 & 1.00 &      3.2 &      8.5 & 146470.2 &         $\cdots$ &         $\cdots$ & $\cdots$ & X14$^{\rm T}$\\
Kepler-351 b  &  4.80 & $\sb{ -4.70}\sp{ +5.70}$ &  3.03 & $\sb{-0.24}\sp{+1.33}$ &  542 &  22.2 &  0.95 & 0.214 & 0.89 &      4.9 & $\cdots$ &    356.0 & 5.6\by$\ttt{33}$ & 3.6\by$\ttt{33}$ &      1.6 & HL14$^{\rm T}$\\
Kepler-351 c  & 11.10 & $\sb{ -7.60}\sp{ +9.90}$ &  3.16 & $\sb{-0.25}\sp{+1.38}$ &  468 &  56.9 &  1.94 & 0.287 & 0.89 &      4.9 & $\cdots$ &    197.9 &         $\cdots$ &         $\cdots$ & $\cdots$ & HL14$^{\rm T}$\\
Kepler-36 b   &  4.46 & $\sb{ -0.27}\sp{ +0.33}$ &  1.48 & $\pm 0.03$             & 1069 &  21.3 &  7.52 & 0.115 & 1.07 &      4.8 &      1.6 &  10334.7 &         $\cdots$ &         $\cdots$ & $\cdots$ & Ca12$^{\rm T}$\\
Kepler-36 c   &  8.10 & $\sb{ -0.46}\sp{ +0.60}$ &  3.68 & $\pm 0.05$             & 1014 &  16.5 &  0.90 & 0.128 & 1.07 &      4.8 &      1.6 &   8348.0 & 2.5\by$\ttt{35}$ & 1.2\by$\ttt{35}$ &      2.1 & Ca12$^{\rm T}$\\
Kepler-385 b  & 12.80 & $\sb{ -6.80}\sp{+11.40}$ &  2.79 & $\sb{-0.35}\sp{+1.62}$ & 1041 &  33.4 &  3.25 & 0.097 & 1.09 &      3.3 & $\cdots$ &   8110.4 &         $\cdots$ &         $\cdots$ & $\cdots$ & HL14$^{\rm T}$\\
Kepler-385 c  & 13.20 & $\sb{ -9.00}\sp{+16.80}$ &  3.10 & $\sb{-0.39}\sp{+1.81}$ &  909 &  35.5 &  2.44 & 0.127 & 1.09 &      3.3 & $\cdots$ &   4731.3 &         $\cdots$ &         $\cdots$ & $\cdots$ & HL14$^{\rm T}$\\
Kepler-396 c  & 17.80 & $\sb{ -1.27}\sp{ +2.86}$ &  5.31 & $\sb{-0.99}\sp{+1.95}$ &  440 &  57.7 &  0.66 & 0.368 & 0.85 &      4.4 &      2.0 &   1032.9 &         $\cdots$ &         $\cdots$ & $\cdots$ & X14$^{\rm T}$\\
Kepler-4 b    & 24.47 & $\pm 3.81$               &  4.01 & $\pm 0.21$             & 1614 &  28.7 &  2.10 & 0.046 & 1.22 &      6.8 &      0.8 &   9535.9 &         $\cdots$ &         $\cdots$ & $\cdots$ & Bo10$^{\rm R}$\\
Kepler-406 b  &  6.36 & $\pm 1.27$               &  1.44 & $\pm 0.03$             & 1452 &  23.1 & 11.83 & 0.036 & 1.07 &      2.1 &      0.2 &    758.8 &         $\cdots$ &         $\cdots$ & $\cdots$ & Mar14$^{\rm R}$\\
Kepler-406 c  &  2.86 & $\pm 1.91$               &  0.85 & $\pm 0.03$             & 1171 &  21.7 & 25.43 & 0.056 & 1.07 &      2.1 &      0.2 &    321.0 &         $\cdots$ &         $\cdots$ & $\cdots$ & Mar14$^{\rm R}$\\
Kepler-454 b  &  6.84 & $\pm 1.40$               &  2.37 & $\pm 0.13$             &  916 &  23.9 &  2.83 & 0.095 & 1.03 &      4.3 &      1.2 &   4051.6 &         $\cdots$ &         $\cdots$ & $\cdots$ & Ge15$^{\rm R}$\\
Kepler-48 b   &  3.94 & $\pm 2.10$               &  1.92 & $\pm 0.10$             & 1024 &  15.2 &  3.07 & 0.053 & 0.88 &      6.2 &      0.3 &    759.5 &         $\cdots$ &         $\cdots$ & $\cdots$ & Mar14$^{\rm R}$\\
Kepler-48 c   & 14.61 & $\pm 2.30$               &  2.71 & $\pm 0.14$             &  809 &  50.5 &  4.05 & 0.085 & 0.88 &      6.2 &      0.3 &    296.8 &         $\cdots$ &         $\cdots$ & $\cdots$ & Mar14$^{\rm R}$\\
Kepler-48 d   &  7.95 & $\pm 4.61$               &  2.04 & $\pm 0.11$             &  492 &  59.9 &  5.14 & 0.230 & 0.88 &      6.2 &      0.3 &     40.7 &         $\cdots$ &         $\cdots$ & $\cdots$ & Mar14$^{\rm R}$\\
Kepler-49 b   &  7.80 & $\sb{ -3.90}\sp{+15.40}$ &  2.72 & $\pm 0.12$             &  627 &  34.7 &  2.14 & 0.060 & 0.55 &      2.9 &      1.5 &   8833.6 &         $\cdots$ &         $\cdots$ & $\cdots$ & X13$^{\rm T}$\\
Kepler-49 c   &  7.90 & $\sb{ -3.90}\sp{+15.60}$ &  2.55 & $\pm 0.13$             &  546 &  43.0 &  2.63 & 0.079 & 0.55 &      2.9 &      1.5 &   5074.4 &         $\cdots$ &         $\cdots$ & $\cdots$ & X13$^{\rm T}$\\
Kepler-51 b   &  2.22 & $\sb{ -0.95}\sp{ +1.59}$ &  7.10 & $\pm 0.30$             &  561 &   4.2 &  0.03 & 0.251 & 1.04 &      3.4 &      3.3 &   3724.8 & 1.8\by$\ttt{39}$ & 1.0\by$\ttt{37}$ &    180.0 & Mas14$^{\rm T}$\\
Kepler-51 c   &  4.13 & $\pm 0.32$               &  9.01 & $\sb{-1.71}\sp{+2.81}$ &  453 &   7.7 &  0.03 & 0.384 & 1.04 &      3.4 &      3.3 &   1596.5 & 5.3\by$\ttt{37}$ & 4.5\by$\ttt{36}$ &     11.8 & Mas14$^{\rm T}$\\
Kepler-51 d   &  7.63 & $\pm 0.95$               &  9.71 & $\pm 0.50$             &  394 &  15.1 &  0.05 & 0.509 & 1.04 &      3.4 &      3.3 &    908.7 & 1.3\by$\ttt{36}$ & 3.8\by$\ttt{35}$ &      3.4 & Mas14$^{\rm T}$\\
Kepler-54 b   & 21.50 & $\sb{ -4.90}\sp{ +5.60}$ &  2.19 & $\pm 0.07$             &  605 & 122.9 & 11.29 & 0.063 & 0.51 &      4.2 &      0.8 &   2727.7 &         $\cdots$ &         $\cdots$ & $\cdots$ & HL14$^{\rm T}$\\
Kepler-54 c   & 19.80 & $\sb{ -4.40}\sp{ +6.60}$ &  1.32 & $\pm 0.08$             &  530 & 214.3 & 47.48 & 0.082 & 0.51 &      4.2 &      0.8 &   1610.1 &         $\cdots$ &         $\cdots$ & $\cdots$ & HL14$^{\rm T}$\\
Kepler-56 b   & 22.25 & $\sb{ -3.50}\sp{ +3.81}$ &  6.52 & $\sb{-0.28}\sp{+0.29}$ & 1496 &  17.3 &  0.44 & 0.103 & 1.32 &      4.4 & $\cdots$ &  18720.8 & 7.8\by$\ttt{35}$ & 4.2\by$\ttt{35}$ &      1.9 & H13$^{\rm T}$\\
Kepler-57 c   &  9.30 & $\sb{ -3.00}\sp{+25.20}$ &  1.55 & $\pm 0.67$             &  705 &  64.5 & 13.77 & 0.092 & 0.76 &      4.7 &      0.8 &   1761.3 &         $\cdots$ &         $\cdots$ & $\cdots$ & X13$^{\rm T}$\\
Kepler-60 b   &  4.20 & $\pm 0.60$               &  1.71 & $\pm 0.13$             & 1177 &  15.8 &  4.63 & 0.073 & 1.04 &      5.1 & $\cdots$ &   8080.2 &         $\cdots$ &         $\cdots$ & $\cdots$ & JH15b$^{\rm T}$\\
Kepler-60 c   &  3.90 & $\pm 0.80$               &  1.90 & $\pm 0.15$             & 1092 &  14.2 &  3.14 & 0.085 & 1.04 &      5.1 & $\cdots$ &   5999.3 &         $\cdots$ &         $\cdots$ & $\cdots$ & JH15b$^{\rm T}$\\
Kepler-60 d   &  4.20 & $\pm 0.80$               &  1.99 & $\pm 0.16$             &  992 &  16.1 &  2.94 & 0.103 & 1.04 &      5.1 & $\cdots$ &   4082.8 &         $\cdots$ &         $\cdots$ & $\cdots$ & JH15b$^{\rm T}$\\
Kepler-65 c   & 26.60 & $\sb{-18.50}\sp{+20.40}$ &  2.61 & $\pm 0.04$             & 1363 &  56.7 &  8.25 & 0.068 & 1.25 &      3.3 &      4.0 &  96951.2 &         $\cdots$ &         $\cdots$ & $\cdots$ & HL14$^{\rm T}$\\
Kepler-68 b   &  8.26 & $\sb{ -2.54}\sp{ +2.22}$ &  2.31 & $\sb{-0.09}\sp{+0.06}$ & 1252 &  21.7 &  3.69 & 0.062 & 1.08 &      6.3 &      0.2 &    441.3 &         $\cdots$ &         $\cdots$ & $\cdots$ & Gi13$^{\rm R}$\\
Kepler-68 c   &  4.77 & $\sb{ -3.50}\sp{ +2.54}$ &  0.95 & $\sb{-0.04}\sp{+0.03}$ & 1033 &  36.7 & 30.30 & 0.091 & 1.08 &      6.3 &      0.2 &    204.7 &         $\cdots$ &         $\cdots$ & $\cdots$ & Gi13$^{\rm R}$\\
Kepler-78 b   &  1.91 & $\pm 0.32$               &  1.18 & $\sb{-0.09}\sp{+0.16}$ & 2217 &   5.5 &  6.43 & 0.009 & 0.76 &      0.8 &      1.7 & 731021.4 &         $\cdots$ &         $\cdots$ & $\cdots$ & Pep13$^{\rm R}$\\
Kepler-79 b   & 10.90 & $\sb{ -6.04}\sp{ +7.31}$ &  3.48 & $\pm 0.07$             &  992 &  24.0 &  1.43 & 0.117 & 1.17 &      3.2 & $\cdots$ &   6920.5 & 7.6\by$\ttt{34}$ & 6.3\by$\ttt{34}$ &      1.2 & JH14$^{\rm T}$\\
Kepler-79 c   &  6.04 & $\sb{ -2.29}\sp{ +1.91}$ &  3.73 & $\pm 0.08$             &  784 &  15.7 &  0.64 & 0.187 & 1.17 &      3.2 & $\cdots$ &   2709.1 & 1.7\by$\ttt{35}$ & 7.4\by$\ttt{34}$ &      2.3 & JH14$^{\rm T}$\\
Kepler-79 d   &  6.04 & $\sb{ -1.59}\sp{ +2.10}$ &  7.17 & $\sb{-0.16}\sp{+0.13}$ &  633 &  10.1 &  0.09 & 0.287 & 1.17 &      3.2 & $\cdots$ &   1150.1 & 1.3\by$\ttt{37}$ & 6.1\by$\ttt{35}$ &     21.3 & JH14$^{\rm T}$\\
Kepler-79 e   &  4.13 & $\sb{ -1.11}\sp{ +1.21}$ &  3.49 & $\pm 0.13$             &  546 &  16.4 &  0.54 & 0.386 & 1.17 &      3.2 & $\cdots$ &    635.8 & 6.8\by$\ttt{34}$ & 2.0\by$\ttt{34}$ &      3.4 & JH14$^{\rm T}$\\
Kepler-80 b   &  5.70 & $\sb{ -4.10}\sp{ +7.40}$ &  2.65 & $\pm 0.11$             &  700 &  23.3 &  1.69 & 0.065 & 0.72 &      2.9 &      1.1 &   3568.7 & 3.6\by$\ttt{34}$ & 1.6\by$\ttt{34}$ &      2.3 & X13$^{\rm T}$\\
Kepler-80 c   &  7.00 & $\sb{ -5.50}\sp{ +8.90}$ &  2.79 & $\pm 0.13$             &  633 &  30.0 &  1.77 & 0.079 & 0.72 &      2.9 &      1.1 &   2391.0 & 1.3\by$\ttt{34}$ & 9.6\by$\ttt{33}$ &      1.4 & X13$^{\rm T}$\\
Kepler-81 b   & 16.40 & $\sb{ -9.90}\sp{+27.00}$ &  2.42 & $\pm 0.38$             &  707 &  72.5 &  6.35 & 0.055 & 0.64 &      3.6 &      2.0 &  17772.7 &         $\cdots$ &         $\cdots$ & $\cdots$ & X13$^{\rm T}$\\
Kepler-81 c   &  4.00 & $\sb{ -2.40}\sp{ +6.70}$ &  2.37 & $\pm 0.37$             &  559 &  22.9 &  1.66 & 0.089 & 0.64 &      3.6 &      2.0 &   6948.7 & 6.6\by$\ttt{34}$ & 3.0\by$\ttt{34}$ &      2.2 & X13$^{\rm T}$\\
Kepler-82 c   & 19.10 & $\sb{ -4.90}\sp{+55.50}$ &  5.35 & $\pm 2.45$             &  500 &  54.0 &  0.69 & 0.257 & 0.85 &      4.7 & $\cdots$ &    265.2 &         $\cdots$ &         $\cdots$ & $\cdots$ & X13$^{\rm T}$\\
\hline
\end{tabular}
\end{table*}

\begin{table*}
\begin{minipage}{\textwidth}
\contcaption{Sub-Neptune Planet Parameters.}
\begin{tabular}{@{\extracolsep{\fill}}lr@{}lr@{}lrrrccrrrccrr}
\hline
Name                   & \mctc{$M\sb{\rm p}$}   & \mctc{$R\sb{\rm p}$}   &
$T\sb{\rm eq}$         & \jep          & \multicolumn{1}{c}{$\rho\sb{\rm p}$} &
$a$                    & $M\sb{\rm s}$          & Age                    &
$\Omega\sb{\rm rot}$   & $F\sb{\rm XUV}$        & $L\sb{\rm hy}$         &
$L\sb{\rm en}$         & $L\sb{\rm hy}$/$L\sb{\rm en}$ &
Ref.$\sp{a,b}$ \\
                       & \mctc{$M\sb{\oplus}$}  & \mctc{$R\sb{\oplus}$}  &
K                      &                        & g\,cm$\sp{-3}$         &
AU                     & $M\sb{\odot}$          & Gyr                    &
$\Omega\sb{\odot}$     & erg\,s\sp{-$1$}\,cm\sp{-$2$}  & s$\sp{-1}$      &
s$\sp{-1}$                    &                             \\
\hline
Kepler-83 c   &  7.50 & $\sb{ -3.60}\sp{+13.80}$ &  2.37 & $\pm 0.35$             &  485 &  49.5 &  3.12 & 0.126 & 0.66 &      3.1 &      1.8 &   2880.2 &         $\cdots$ &         $\cdots$ & $\cdots$ & X13$^{\rm T}$\\
Kepler-84 b   & 21.20 & $\sb{-13.80}\sp{+32.00}$ &  2.23 & $\pm 0.10$             & 1092 &  65.9 & 10.50 & 0.083 & 1.00 &      9.6 &      1.4 &  11252.1 &         $\cdots$ &         $\cdots$ & $\cdots$ & X13$^{\rm T}$\\
Kepler-85 b   & 15.30 & $\sb{-10.80}\sp{+22.10}$ &  1.97 & $\pm 0.10$             &  884 &  66.4 & 10.95 & 0.078 & 0.92 &      4.0 &      1.2 &   4726.9 &         $\cdots$ &         $\cdots$ & $\cdots$ & X13$^{\rm T}$\\
Kepler-85 c   & 24.00 & $\sb{-16.70}\sp{+34.40}$ &  2.18 & $\pm 0.10$             &  771 & 108.3 & 12.83 & 0.103 & 0.92 &      4.0 &      1.2 &   2738.9 &         $\cdots$ &         $\cdots$ & $\cdots$ & X13$^{\rm T}$\\
Kepler-87 c   &  6.40 & $\pm 0.80$               &  6.15 & $\pm 0.09$             &  444 &  17.7 &  0.15 & 0.671 & 1.10 &      7.2 &      1.3 &    178.9 & 9.6\by$\ttt{34}$ & 2.5\by$\ttt{34}$ &      3.8 & O14$^{\rm T}$\\
Kepler-88 b   &  8.58 & $\pm 2.54$               &  3.78 & $\sb{-0.36}\sp{+0.39}$ &  801 &  21.5 &  0.88 & 0.095 & 0.96 &      2.2 &      1.1 &   2566.9 & 6.2\by$\ttt{34}$ & 2.2\by$\ttt{34}$ &      2.8 & N13$^{\rm T}$\\
Kepler-89 b   & 10.49 & $\pm 4.45$               &  1.72 & $\pm 0.16$             & 1624 &  28.5 & 11.43 & 0.051 & 1.28 &      3.3 &      2.6 &  78217.2 &         $\cdots$ &         $\cdots$ & $\cdots$ & W13$^{\rm R}$\\
Kepler-89 c   &  9.40 & $\sb{ -2.10}\sp{ +2.40}$ &  4.41 & $\pm 0.42$             & 1154 &  14.0 &  0.60 & 0.101 & 1.28 &      3.3 &      2.6 &  20126.4 & 7.7\by$\ttt{35}$ & 4.3\by$\ttt{35}$ &      1.8 & Mas13$^{\rm T}$\\
Kepler-89 e   & 13.00 & $\sb{ -2.10}\sp{ +2.50}$ &  6.70 & $\pm 0.63$             &  665 &  22.1 &  0.24 & 0.305 & 1.28 &      3.3 &      2.6 &   2227.5 & 1.7\by$\ttt{35}$ & 7.7\by$\ttt{34}$ &      2.2 & Mas13$^{\rm T}$\\
Kepler-92 c   &  6.04 & $\pm 1.91$               &  2.60 & $\pm 0.08$             &  856 &  20.5 &  1.89 & 0.186 & 1.21 &      5.9 &      1.7 &   3067.7 & 4.3\by$\ttt{34}$ & 1.6\by$\ttt{34}$ &      2.8 & X14$^{\rm T}$\\
Kepler-93 b   &  4.02 & $\pm 0.68$               &  1.48 & $\pm 0.02$             & 1138 &  18.1 &  6.82 & 0.053 & 0.91 &      5.2 &      0.3 &    942.4 &         $\cdots$ &         $\cdots$ & $\cdots$ & Dr15$^{\rm R}$\\
Kepler-94 b   & 10.81 & $\pm 1.27$               &  3.51 & $\pm 0.15$             & 1095 &  21.3 &  1.38 & 0.034 & 0.81 &      2.5 &      0.4 &   1815.1 & 3.5\by$\ttt{34}$ & 1.2\by$\ttt{34}$ &      2.9 & Mar14$^{\rm R}$\\
Kepler-95 b   & 13.03 & $\pm 2.86$               &  3.42 & $\pm 0.09$             & 1019 &  28.3 &  1.79 & 0.102 & 1.08 &      8.7 &      0.3 &    281.0 & 2.6\by$\ttt{33}$ & 2.8\by$\ttt{33}$ &      0.9 & Mar14$^{\rm R}$\\
Kepler-96 b   &  8.58 & $\pm 3.50$               &  2.67 & $\pm 0.22$             &  782 &  31.2 &  2.48 & 0.125 & 1.00 &      3.7 &      0.3 &    128.8 &         $\cdots$ &         $\cdots$ & $\cdots$ & Mar14$^{\rm R}$\\
Kepler-97 b   &  3.50 & $\pm 1.91$               &  1.48 & $\pm 0.13$             & 1451 &  12.3 &  5.93 & 0.036 & 0.94 &      4.6 &      0.3 &   1966.4 &         $\cdots$ &         $\cdots$ & $\cdots$ & Mar14$^{\rm R}$\\
Kepler-98 b   &  3.50 & $\pm 1.59$               &  2.00 & $\pm 0.22$             & 1743 &   7.6 &  2.42 & 0.026 & 0.99 &      8.3 &      0.2 &   2774.0 &         $\cdots$ &         $\cdots$ & $\cdots$ & Mar14$^{\rm R}$\\
Kepler-99 b   &  6.15 & $\pm 1.30$               &  1.48 & $\pm 0.08$             &  880 &  35.8 & 10.46 & 0.050 & 0.79 &      4.3 &      0.4 &    883.3 &         $\cdots$ &         $\cdots$ & $\cdots$ & Mar14$^{\rm R}$\\
WASP-47 d     & 15.20 & $\pm 7.00$               &  3.64 & $\pm 0.13$             &  983 &  32.2 &  1.74 & 0.086 & 1.04 & $\cdots$ &      1.4 &   6812.1 & 3.6\by$\ttt{34}$ & 2.8\by$\ttt{34}$ &      1.3 & Be15$^{\rm T}$\\
\hline
\end{tabular}
\vspace{-0.5cm}
\footnotetext[1]{References for Planetary Masses and Radii.~
 Alm15: \citet{AlmenaraEtal2015aaK2-3},
 Alo14: \citet{AlonsoEtal2014aaCorot24c},
 BT15: \citet{Berta-ThompsonEtal2015natHD1132b},
 Ba14: \citet{BarrosEtal2014aaCorot7b},
 Ba15: \citet{BarrosEtal2015mnrasK2-19},
 Bak10: \citet{BakosEtal2010apjHATP11b},
 Be15: \citet{BeckerEtal2015apjTTVwasp47},
 Bi14: \citet{BiddleEtal2014mnrasRVgj3470},
 Bo10: \citet{BoruckiEtal2010apjKepler4b},
 Ca12: \citet{CarterEtal2012sciTTVkepler36},
 Co11: \citet{CochranEtal2011apjsTTV+RVkepler18},
 De11: \citet{DemoryEtal2011aa55cnceTransit},
 Dr15: \citet{DressingEtal2015apjTTVkepler93},
 E16: \citet{EspinozaEtal2016apjBD20594b},
 Ga12: \citet{GautierEtal2012apjKepler20},
 Ge15: \citet{GettelEtal2015arxivRVkepler454},
 Gi13: \citet{GillilandEtal2013apjTTVkepler68},
 H13: \citet{HuberEtal2013sciTTVkepler56},
 HL14: \citet{HaddenLithwick2014apj139KeplerTTV},
 HL16: \citet{HaddenLithwick2016apjTTVkepler33},
 Ha11: \citet{HartmanEtal2011apjHATP26b},
 Ha13: \citet{HarpsoeEtal2013aaGJ1214},
 JH14: \citet{Jontof-HutterEtal2014apjTTVkepler79},
 JH15a: \citet{Jontof-HutterEtal2015natTTVkepler138},
 JH15b: \citet{Jontof-HutterEtal2015axivTTVkeplers},
 K14: \citet{KippingEtal2014apjTTVkepler231},
 L13: \citet{LissauerEtal2013apjTTVkepler11},
 Mac14: \citet{MaciejewskiEtal2014acaRVgj436},
 Mar14: \citet{MarcyEtal2014apjsRVkeplers},
 Mas13: \citet{MasudaEtal2013apjTTVkepler89},
 Mas14: \citet{Masuda2014apjTTVkepler51},
 Mot15: \citet{MotalebiEtal2015aaRVhd219134},
 Mou14: \citet{MoutouEtal2014mnrasCortot22b},
 N13: \citet{NesvornyEtal2013apjTTVkepler88},
 O14: \citet{OfirEtal2014aaTTVkepler87},
 Pep13: \citet{PepeEtal2013natKepler78b},
 Pet15: \citet{PetiguraEtal2015arxivRVepic2037},
 S14: \citet{SchmittEtal2014apjTTVkepler289},
 SO12: \citet{Sanchis-OjedaEtal2012natTTVkepler30},
 V15: \citet{VanderburgEtal2015apjHIP116454b},
 VG14: \citet{vanGrootelEtal2014apjHD97658b},
 W13: \citet{WeissEtal2013apjRVkepler89},
 W16: \citet{WeissEtal2016Kepler10},
 X13: \citet{Xie2013apjsTTV},
 X14: \citet{Xie2014apjsTTV}.}
\footnotetext[2]{The `R' and `T' superscripts indicate RV- and TTV-estimated mass, respectively.}
\end{minipage}
\end{table*}
\normalsize
}

\bsp	
\label{lastpage}
\end{document}